\documentclass[aps,prl,twocolumn]{revtex4-1}
\usepackage{diagbox}
\usepackage{amsfonts}
\usepackage{amsmath}
\usepackage{amsthm}
\usepackage{indentfirst}
\usepackage{microtype}
\usepackage[squaren]{SIunits}
\usepackage{booktabs}
\usepackage{graphicx}
\usepackage{multirow}
\usepackage{wallpaper}
\usepackage{soul}

\newcommand{\beq}{\begin{equation}}
\newcommand{\eeq}{\end{equation}}
\newcommand{\bea}{\begin{eqnarray}}
\newcommand{\eea}{\end{eqnarray}}

\usepackage{braket}
\usepackage{caption}
\usepackage{subcaption}
 \makeatletter
 \g@addto@macro\normalsize{
   \setlength\abovedisplayskip{15pt}
   \setlength\belowdisplayskip{15pt}
   \setlength\abovedisplayshortskip{15pt}
   \setlength\belowdisplayshortskip{15pt}
 }
\newcommand{\CP}[1]{{\color{black}{#1}}}

\makeatother

\begin{document}
\title{Energy gap closure of crystalline molecular hydrogen with pressure}
\author{Vitaly Gorelov$^1$, Markus Holzmann$^{2,3}$,  David M. Ceperley$^4$, Carlo Pierleoni$^{1,5,*}$ }

\affiliation{
$^1$Maison de la Simulation, CEA-CNRS-UPS-UVSQ, Universit\'e Paris-Saclay, 91191 Gif-sur-Yvette, France\\
$^2$ Univ. Grenoble Alpes, CNRS, LPMMC, 38000 Grenoble, France \\
$^3$ Institut Laue-Langevin, BP 156, F-38042 Grenoble Cedex 9, France\\
$^4$Department of Physics, University of Illinois Urbana-Champaign, USA\\ 
$^5$ Department of Physical and Chemical Sciences, University
of L'Aquila, Via Vetoio 10, I-67010 L'Aquila, Italy\\
}

\date{\today}
\begin{abstract}
We study the gap closure with pressure of crystalline molecular hydrogen.  The gaps are obtained from grand-canonical Quantum Monte Carlo methods properly extended to quantum and thermal crystals, simulated by Coupled Electron Ion Monte Carlo. Nuclear zero point effects cause a large reduction in the gap ($\sim 2eV$).  \CP{Depending on the structure,} the fundamental indirect gap closes \CP{between 380GPa and} 530GPa for ideal crystals and 330-380GPa for quantum crystals.  Beyond this pressure the system enters into a bad metal phase where the density of states at the Fermi level increases with pressure up to $\sim$450\CP{-500} GPa when the direct gap closes. Our work partially supports the interpretation of recent experiments in high pressure hydrogen.
\end{abstract}

\maketitle
The metallization of crystalline hydrogen under pressure has attracted considerable attention over the last century. Predicted to be stable in an atomic bcc lattice around 25GPa, the mechanism for molecular dissociation was first discussed by Wigner and Huntington \cite{Wigner1935}. The search for its metallization has driven high pressure research until the recent \cite{Dias2017}, still debated \cite{Goncharov2017,Loubeyre2017,Liu2017,Silvera2017a}, observation of reflective samples at 495GPa in a Diamond Anvil Cell (DAC) apparatus. 
Even though it is the simplest element and H$_2$ the simplest homonuclear molecule in nature, 
the study of hydrogen under extreme conditions has uncovered rich and unexpected physics \cite{Mao1994,McMahon2012a,Nellis2006}.

The mechanism by which the insulating crystal transforms into a conducting crystal is still unclear. Experiments have difficulty in determining the crystalline structure and its evolution with pressure because of the low cross section to X-rays \cite{Loubeyre1996,Ji2019,Dubrovinsky2019} and the  small volume of the samples for neutron scattering. Structural information are obtained indirectly through vibrational spectroscopy while electronic structure is probed by optical measurements \cite{Silvera2018}. Direct measurements of static conductivity in the DAC remain inconclusive \cite{Eremets2011,Nellis2012,Goncharov2012,Goncharov2013,Eremets2016,Eremets2017}. 
A complex phase diagram comprising up to at least four different molecular phases (from I to IV) with different vibrational spectra has been traced experimentally \cite{McMahon2012a}. Recent experiments  \cite{Dias2016,Dias2017,Eremets2017,Loubeyre2019,Dias2019} searched for  metallization at low temperature ($\leq$ 100K) while raising pressure in phase III. Considerable attention has also been paid to the higher temperature phase IV since its discovery \cite{Eremets2011,Howie2012,Howie2012b,Zha2012,Loubeyre2013,Howie2015}. 
The emerging picture is that the transparent insulating molecular phase III transforms into a strongly absorbing (in the visible) molecular phase at $\sim$ 350-360GPa with different IR frequencies, \CP{first named phase V\cite{Eremets2016}} and later $H_2$-PRE or phase VI \cite{Silvera2018,Dias2019}, with semiconducting characteristics \cite{Eremets2019}. Hydrogen finally reaches a metallic phase with the observation of reflective samples at $\sim$495GPa\cite{Dias2017}, although disagreement concerning the pressure scale still remains \cite{Eremets2016a,Loubeyre2017,Silvera2018}. New synchrotron infrared spectroscopy measurements \cite{Loubeyre2019} report a reversible collapse of the IR transmission spectrum at 427GPa, interpreted as a first order transition to the metallic state, an interpretation criticised in \cite{Silvera2019}. 

In this paper we investigate the closure of the electronic gap of candidate structures for phase III (Cmca-12 and C2/c-24) and phase IV (Pc48)\cite{Pickard2007,Pickard2012}  within a Quantum Monte Carlo (QMC) framework \cite{Yang2019}.
For ideal structures, the fundamental gap decreases with pressure from $\sim$ 3-3.5 eV at $\sim$250GPa to a vanishing value $\sim$380GPa in the Cmca12 structure and $\sim$530GPa in the C2/c-24 structure. 
Using Coupled Electron-Ion Monte Carlo (CEIMC) calculations, we then include
Zero Point Motion (ZPM) and finite temperature
effects of the nuclei within a first principles, non-perturbative Path Integral approach.
Extending the grand canonical method \cite{Yang2019} to 
quantum crystal at finite temperature, we 
observe a strong gap reduction of $\sim$ 2eV due to nuclear quantum effects (NQE) while temperature effects below 300K are minor. 
At 200K the fundamental indirect gap closes  $\sim$330GPa for Cmca-12 and $\sim$380GPa for C2/c-24. Raising the temperature of C2/c-24 to 290K reduces the closure pressure to 340GPa while decreasing it to 100K does not give any noticeable effect.
\CP{For both structures the direct gap, as obtained by unfolding of the supercell bands \cite{SupMat}, remains open up to $\sim$470-500GPa. Values for the C2/c-24 structure are in agreement with recent experimental data \cite{Loubeyre2019}, although we cannot discuss the experimentally observed sudden closure at 427GPa.}
Our new method for calculating energy gaps allows us to benchmark DFT functionals not only for thermodynamics and structural properties, but also for excitation energies, important for predicting optical properties.

{\bf Method.} The primary information for theoretical investigations of solids are the crystalline structures.
Candidates structures for high pressure phases have been obtained by ab initio Random Structural Search methods \cite{Pickard2007,Pickard2012,Pickard2012erratum,Monserrat2018}. 
 For phase III we consider C2/c-24 and Cmca-12, which are among the lowest free energy structures in ground state QMC calculations assuming harmonic phonons corrections (with DFT-PBE) \cite{McMinis2015b,Azadi2014}. For Phase IV we consider only Pc48, since the recently proposed Pca21 structure \cite{Monserrat2018} is found to be rather similar to Pc48 after geometry relaxation.
We first consider ideal crystal structures (protons fixed at lattice sites) relaxed at constant pressure with the DFT-vdW-DF functional. 
Quantum crystals, with protons represented by path integrals at finite temperature, are addressed with CEIMC at constant volume\footnote{We have checked that the stress tensor in the constant volume CEIMC run remains diagonal with same diagonal elements within our statistical noise.}.
All systems considered have 96 protons in nearly cubic supercells. Optimized Slater-Jastrow-Backflow trial wave functions have been used for the CEIMC calculations \cite{Pierleoni2016}; details of the CEIMC simulations are reported in Ref.\cite{Rillo2018}.
Averages over ionic positions for gaps are obtained using 40 statistically independent configurations from the CEIMC trajectories.

For a given fixed nuclear configuration, the fundamental energy gap is obtained by considering systems with a variable number of electrons $n\in [-6,6]$ where $n=N_e-N_p$. For each system we perform \CP{Reptation Quantum Monte Carlo (RQMC)} calculations with imaginary-time projection $t=$2.00 Ha$^{-1}$ and time step $\tau=$0.01 Ha$^{-1}$ for up to $6\times 6\times 6$ Monkhorst-Pack grid of twists. We check that those values are adequate for converging the band gaps within our resolution. 
The fundamental gap is obtained from \CP{grand-canonical twist-averaged boundary conditions (GCTABC) RQMC} and corrected for finite size effects in leading and next-to-leading order \cite{Yang2019}. 

\begin{figure}
  \includegraphics[width=\columnwidth]{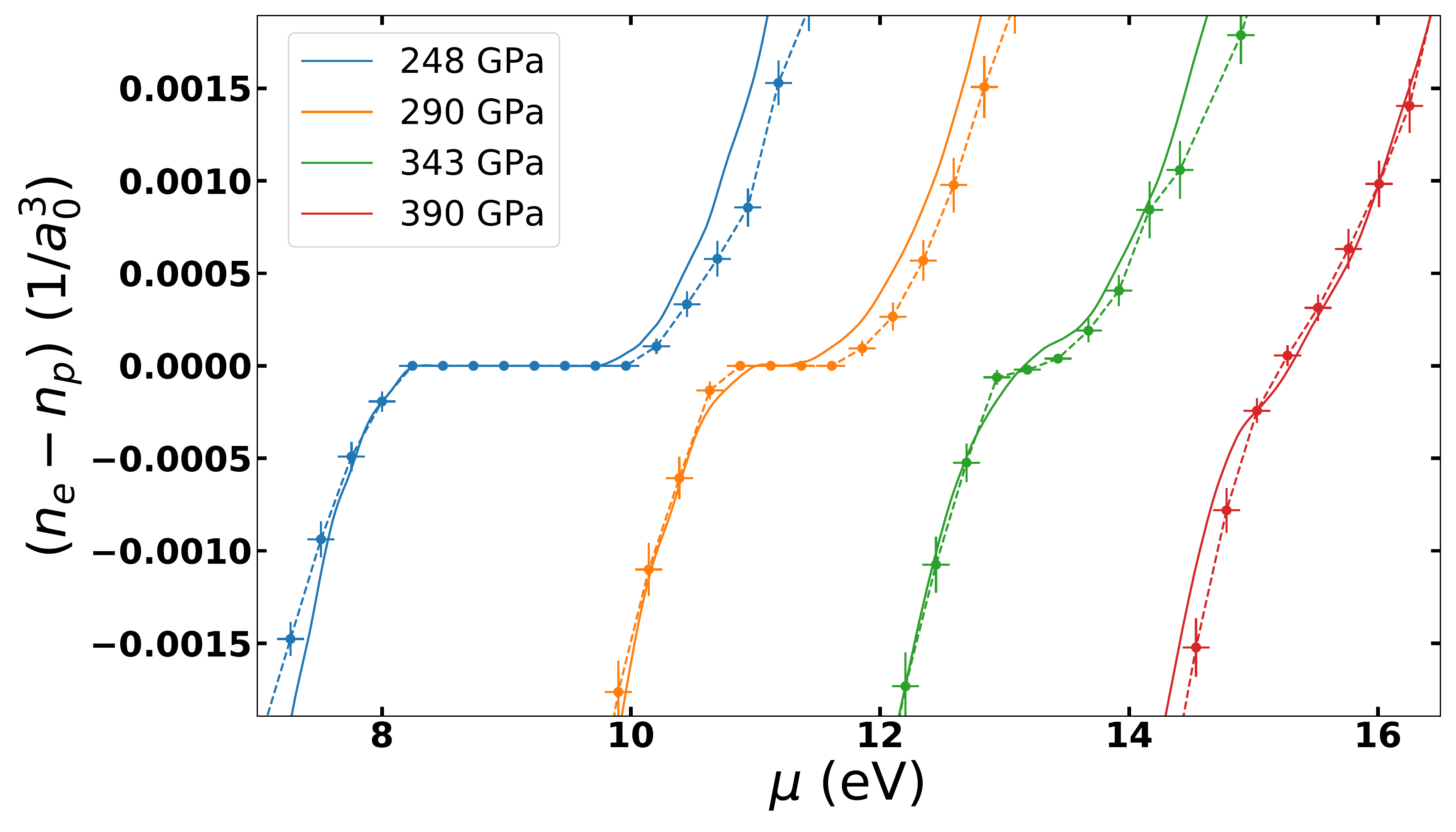}
  \caption{\small{Integrated density of states for C2/c-24 quantum crystals at 200K from GCTABC-RQMC (points) and HSE (smooth lines) at various pressures. }}
  \label{fig:DOS}
\end{figure}

Extending calculations of the fundamental gaps to quantum crystals, 
the trace over nuclear degrees of freedom must be taken with care.
In the semiclassical approximation \cite{SupMat}, the fundamental gap is the smallest electronic excitation energy that occurs from quantum or thermal fluctuations of the lattice. 
Strictly speaking this gap is always closed, since the probability of a proton configuration with a metallic character is never exactly zero.
For dense molecular hydrogen phonon energies are $\sim 0.1-0.5$ eV \cite{Pickard2012}. ZPM dominates for $T\leq1000$K, so the semi-classical approach is not appropriate. 
Electronic energies should be averaged over the nuclear configurations \CP{according to their thermal distribution.}
The gap will be given by the minimum of the average excitation energies, always larger than the semiclassical gap.
Figure~\ref{fig:DOS} illustrates typical results for the integrated density of states as a function of (electronic) chemical potential. The gap of the quantum crystal can be directly read off from the width of the incompressible region. 
More details are given in \cite{SupMat}. 

\begin{figure}
  \includegraphics[width=\columnwidth]{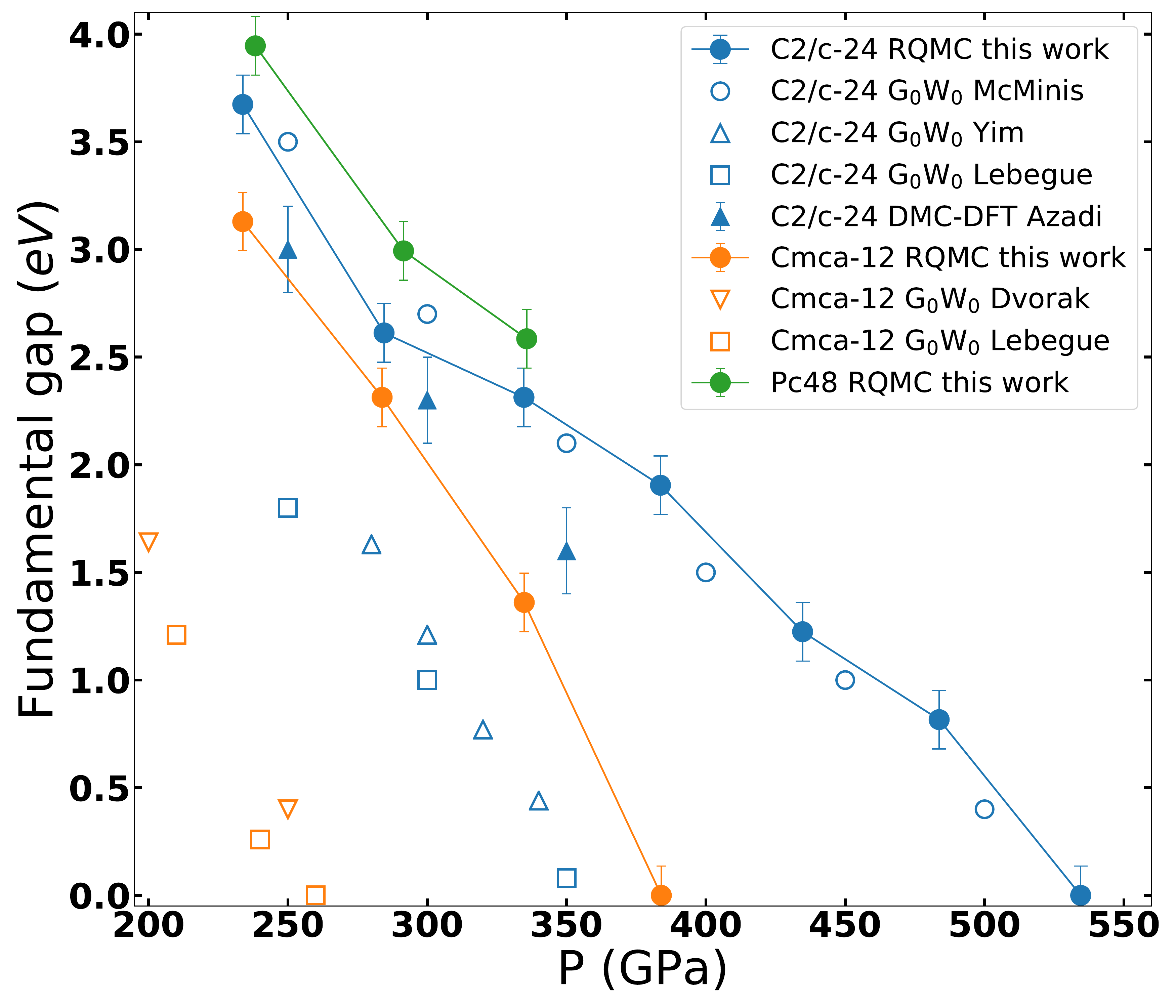}
  \caption{\small{Fundamental energy gap for ideal crystals. This work (closed circles): C2/c-24 (blue), Cmca-12 (orange) and Pc48 (green), open GW results for C2/c-24 (open blue circles\cite{McMinis2015b}). These structures were optimized with vdW-DF functional. QMC for C2/c-24 optimized with the BLYP from ref.\cite{Azadi2017} (closed blues triangles). GW results from Refs. \cite{Lebegue2012,Yim,Dvorak2014} for C2/c-24 (blue) and Cmca-12 (orange) optimized with the PBE functional. Note that pressures from RQMC are 10-15GPa lower than the nominal optimization pressure.}}
  \label{fig:IdealGap}
\end{figure}    

{\bf Results.} 
Figure \ref{fig:IdealGap} shows estimates of the fundamental gap for ideal crystals versus pressure. 
The gap decreases with pressure in a similar fashion for all structures:
Cmca-12 has the smallest gap, followed by C2/c-24 and by Pc48. We find reasonable agreement with the QMC estimates of Ref.~\cite{Azadi2017} \footnote{The observed small difference, in particular at the higher pressure, is probably due to the different XC approximation used for geometry optimization, vdW-DF in our case, BLYP in Ref. \cite{Azadi2017} and different size extrapolation.}. 
References~\cite{Lebegue2012,Yim,Dvorak2014} report smaller values of the gap based on GW. We believe this disagreement is primarily due to the lattice geometry that has been optimized at constant pressure with PBE in Refs \cite{Lebegue2012,Yim,Dvorak2014} and with vdW-DF in the present work.
It has been previously observed that PBE optimized geometries has longer H$_2$ bonds and
smaller gap values \CP{at DFT level}\cite{Morales2013,Clay2014}. \CP{This propagates into $G_0W_0$}. Indeed, GW results from structures optimized with vdw-DF \cite{McMinis2015b} are in excellent agreement with our predictions. 

\begin{figure}
  \includegraphics[width=\columnwidth]{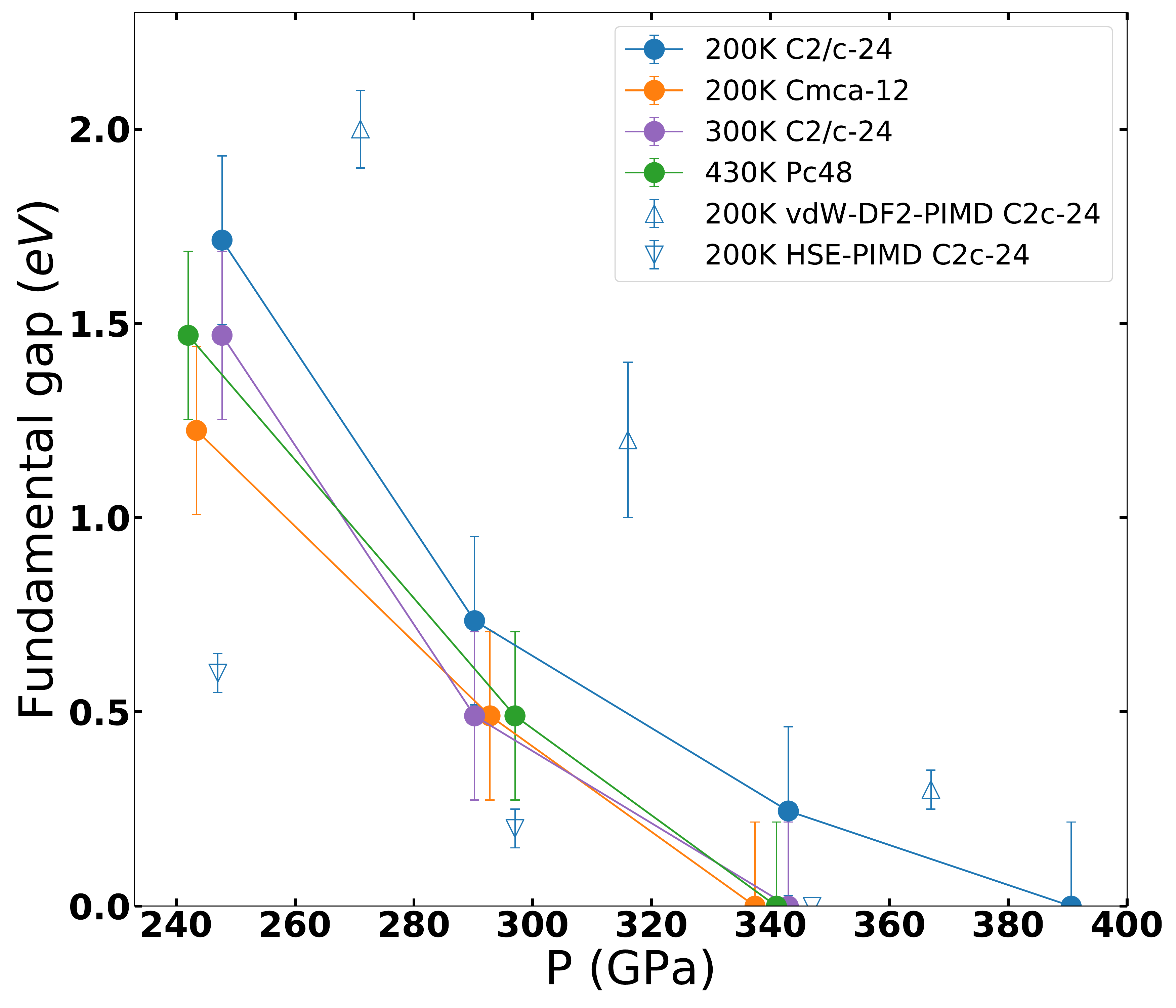}
  \caption{\small{The fundamental gap of quantum crystals at finite temperature. Closed circles indicate results from this work, for the three structures at various temperature as detailed in the legend. PIMD-DFT results at 200K are obtained with two different XC approximations, namely HSE (downward open triangles) and vdW-DF2(upward open triangles) and the semiclassical averaging are reported for comparison \cite{Morales2013}.}}
  \label{fig:QuantumGap}
\end{figure}    
Values of the fundamental gap from \CP{GCTABC} for quantum crystals at various temperatures and pressures are shown in Fig.~\ref{fig:QuantumGap}: they are smaller by $\sim$2eV with respect to the ideal crystal.
ZPM is almost entirely responsible for this reduction.
Note that the gap hardly changes from 300K to 200K within our estimated errors. 
Similar to ideal crystals, Cmca-12 gap is smaller than C2/c-24 gap at T=200K, the former closing  $\sim$340GPa while the latter at higher pressures $\sim$380GPa. As for the Pc-48 structure at T=430K (phase IV) the gap is slightly below values for C2/c-24 at 200K. Our results show that the electronic gap is fairly independent of the specific crystalline structure of the molecular quantum crystals.
We also report gap values for C2/c-24 at T=200K from Path Integral Molecular Dynamics (PIMD)\cite{Morales2013} with two different DFT functionals, namely HSE \cite{Heyd2005} and vdW-DF2 \cite{Lee2010}.
As vdW-DF2 underestimates the molecular bond lengths of the ideal crystalline structure \cite{Clay2014}, its PIMD configurations are expected to bias the electronic gap towards larger values.
Our results do not agree with predictions of Ref.~\cite{Azadi2018} (not shown) yielding a metallic state 
for C2/c-24 at 300GPa and 300K, \CP{and predict substantially larger gap reduction for C2/c-24 quantum
crystals than Ref.~\cite{Li2013}.} However, those works are based on less controlled assumptions such as using ``scissor corrected'' BLYP band structure and an ad hoc procedure for including nuclear motion.
\begin{figure}
  \includegraphics[width=\columnwidth]{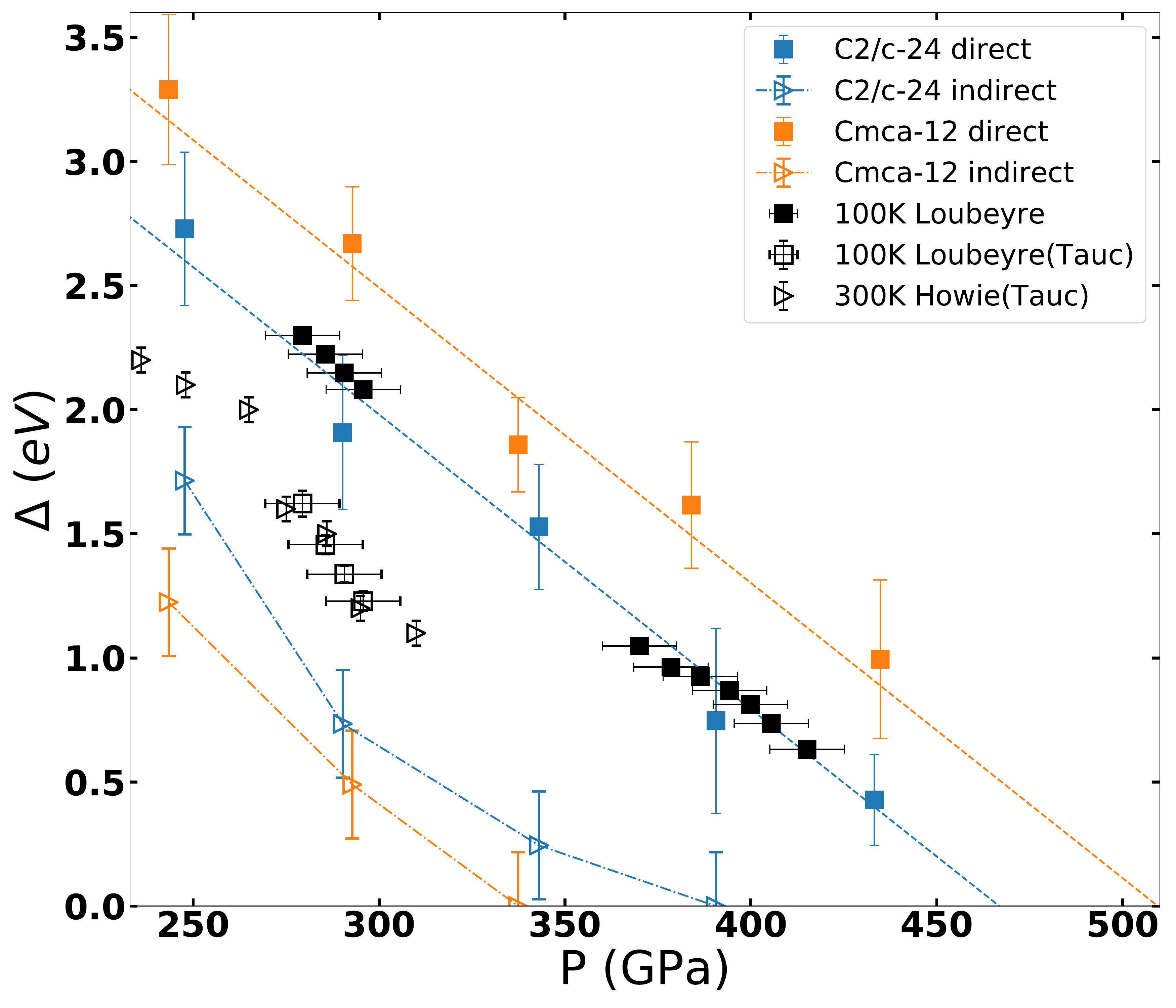}
  \caption{{\small Direct (closed symbols) and indirect (open symbols) gaps of quantum crystals. GCTABC-RQMC at T=200K: C2/c-24 indirect (blue triangles), direct (blues squares); Cmca-12 indirect (orange triangles), direct (closed squares). Experiments: indirect gap from the Tauc analysis at 100K (phase III), (black squares)\cite{Loubeyre2002}, and at 300K (phase IV), (black triangles) \cite{Howie2012b,Goncharov2013}; direct gap at 100K (black squares) \cite{Loubeyre2002,Loubeyre2019}. 
}
  }
  \label{fig:BGabs}
\end{figure}

For all structures considered the observed fundamental gap is indirect. Estimate of the direct gap can be obtained 
\CP{by unfolding the band structure of the supercell \cite{SupMat}.
Fig.~\ref{fig:BGabs} shows the direct gap for both C2/c-24 and Cmca-12 structures. While for the indirect gap Cmca-12 is always lower than C2/c-24, the direct gap is systematically larger. The difference between direct and indirect gap is of $\sim 1eV$ for C2/c-24, and of $\sim 2eV$ for Cmca-12. Closure of the direct gaps, obtained by linear extrapolation, occurs $\sim 450GPa$ in C2/c-24 and $\sim 500GPa$ in Cmca-12. Hence for both structures we observe an intermediate pressure region where the fundamental indirect gap is closed but the direct vertical gap remains open and decreases linearly with pressure.} 
In this region, we expect the density of states around the Fermi level to increases progressively with pressure, as qualitatively reported in Ref.~\cite{Rillo2018}.
This indicates the formation of a bad metal with properties similar to a semi-metal upon closure of the indirect gap, a scenario strongly supporting the recently proposed experimental picture \cite{Eremets2019}(see also refs. \cite{Silvera2018,Dias2019}). 
The non-vanishing direct gap naturally explains the reported observation of absorbing (black) hydrogen around 320-360 GPa (depending on the experimental pressure scale) \cite{Loubeyre2002}.

Fig.~\ref{fig:BGabs} also shows the experimental estimates of both indirect and direct gaps from optical absorption. \CP{Measuring indirect gaps is difficult in hydrogen since samples are very thin and the optical signal from phonon-assisted absorption is too low to be detected \cite{Goncharov2012,Eremets2017}}.
The indirect gap value has been extracted from a Tauc analysis of the absorption profiles at 300K (Phase IV) \cite{Howie2012b,Goncharov2013} and 100K (Phase III) \cite{Loubeyre2002,Zha2012} \CP{assuming the low energy absorption spectrum can be reliably extrapolated to zero energy.} \footnote{We have re-analyzed the spectra of ref. \cite{Loubeyre2002} to extract the value of the indirect gap from a Tauc plot \cite{Tauc1966}, as was performed in ref. \cite{Goncharov2013} for the data from ref. \cite{Howie2012b}. Details are given in the \cite{SupMat}}. \CP{Conversely the direct gap at 100K (phase III) has been associated with the absorption edge at lower pressure \cite{Loubeyre2002} or with full absorption at higher pressure \cite{Loubeyre2019} and corresponds roughly to the energy where the absorption coefficient equals 30000cm$^{-1}$}. 
\CP{The direct gap of C2/c-24 structure is in agreement with the experimental data up to 425GPa, where experiments report a collapse of the gap value ascribed to the metallization transition\cite{Loubeyre2019}. Our results do not allow to predict this transition, but rule out C2/c-24 and Cmca-12 for this new metallic phase. \footnote{Our estimates of the direct gap could be biased by $\sim 0.3eV$ due to the discreteness of our twist grid. Correcting for this bias will place the experimental data in between the C2/c-24 and Cmca-12 predictions.}
For the indirect gap we predict  $\sim 0.3-0.5eV$ smaller values than in experiments.
However, the Tauc analysis of Refs~\cite{Howie2012b,Goncharov2013,Loubeyre2002} does not consider the energy of the emitted or absorbed phonons, which should be comparable to the observed discrepancy. \CP{However, excitonic effects
and exciton-phonon coupling, neglected within the present approach, need to be addressed for this level of precision.}
 In agreement with our findings, the experimental indirect gap depends little on both temperature and structure \footnote{Note that the pressure values of Ref.~\cite{Loubeyre2002} have been recently corrected \cite{Loubeyre2019}}.}

\begin{figure}
  \includegraphics[width=\columnwidth]{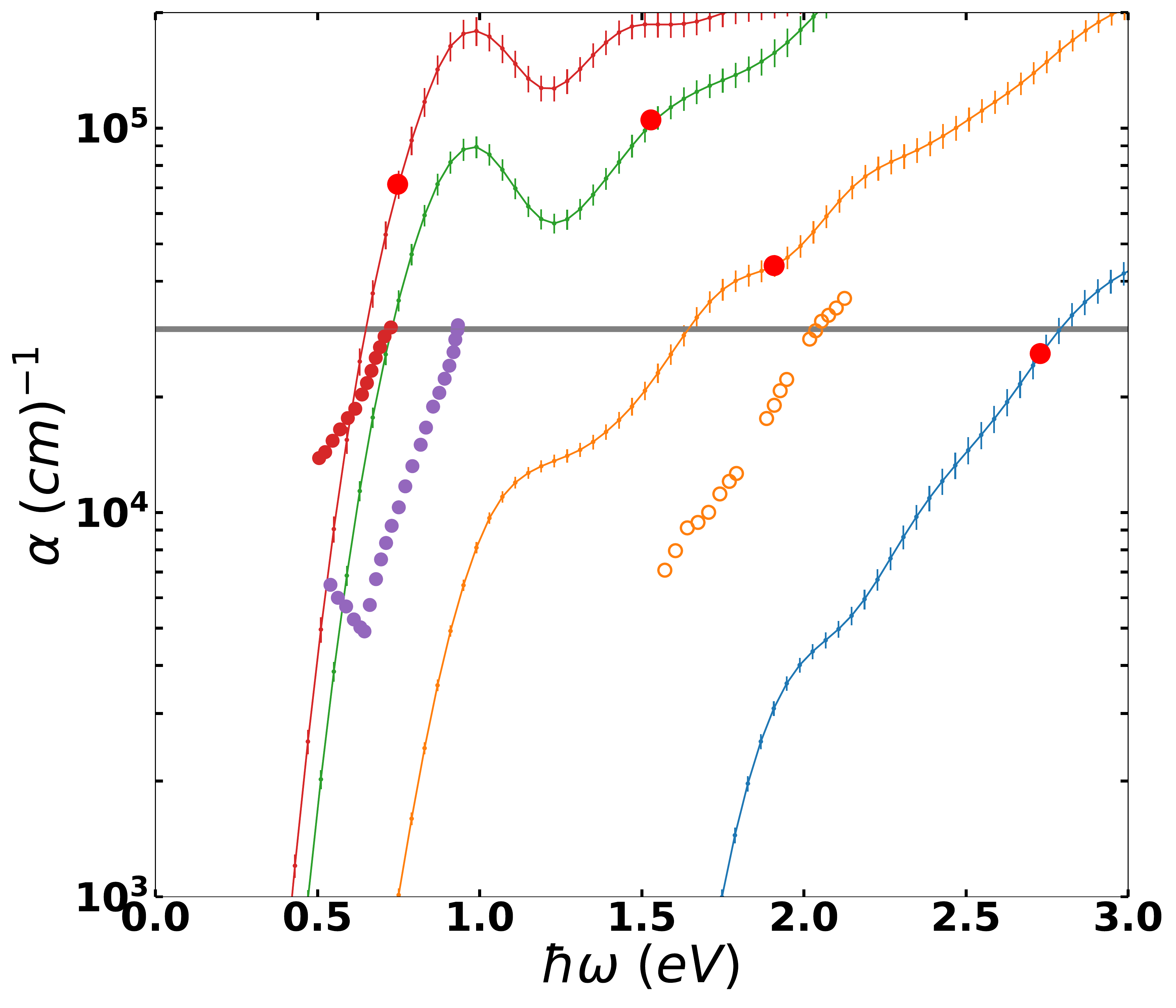}
  \caption{\small{Absorption spectra from HSE band structure for C2/c-24 quantum crystals (solid lines) and comparison with the available experimental profiles (opened and filled circles). The spectra from HSE has been shifted in energy by an amount equal to the difference between QMC and HSE direct gap. The reported pressure are as in figure \ref{fig:DOS} (see the colors). The red dots indicate the location in energy of the direct gap of figure \ref{fig:BGabs}. Experimental pressures are: 296GPa - open orange circles \cite{Loubeyre2002} (corrected by 20 GPa\cite{Loubeyre2019}), 386GPa - magenta filled circles and 406GPa - red filled circles \cite{Loubeyre2019}) }}
  \label{fig:absDFT}
\end{figure}

\CP{Next we explore optical properties computed using the Kubo-Greenwood (KG) framework with Kohn-Sham (KS) orbitals}. To reduce the bias of the underlying DFT functional, we have benchmarked several XC approximations to reproduce the behavior of the QMC density of states close to the gap.
In Fig.~\ref{fig:DOS} for C2/c-24 at 200K, we compare the electronic excess density, $n_e-n_p$, as a function of electronic chemical potential, $\mu$, from QMC and from DFT-HSE \footnote{This quantity is closely related to the integrated density of states.}. The observed plateau at $n_e-n_p=0$ is the signature of the indirect gap. 
Deviations from the plateau on both sides characterize the density of states of the valence and conduction band close to the band edges.
As shown in Fig.~\ref{fig:DOS} the HSE approximation provides slightly smaller values of the fundamental gap and reproduces reasonably well the integrated density of states from \CP{GCTABC} around the Fermi energy (more details are in \cite{SupMat}). We therefore employed HSE to compute optical properties exploiting the KGEC code \cite{Calderin2017a} in the QuantumEspresso suite \cite{Giannozzi2017}. For thermal and quantum crystals considered here, 
the William-Lax (WL) semiclassical (SC) approximation \cite{Williams1951,Lax1952,Patrick2014,Zacharias2015,Zacharias2016} is not appropriate as already discussed. 
Instead of a joint density of states
based on excitation energies for each nuclear configuration entering the WL expression, we have used the corresponding one based on electronic energies averaged over ionic ZPM, more appropriate for low temperatures \cite{SupMat}.
In Fig.~\ref{fig:absDFT} we compare the absorption profiles for C2/c-24 at T=200K and different pressures \footnote{To partially correct for HSE inaccuracy, we shifted the energy scale by the difference between the QMC and HSE gap.} to experimental profiles from Refs \cite{Loubeyre2002, Loubeyre2019} at T=100K. We observe a higher absorption than in experiments at comparable pressure, which cannot be explained by the temperature difference. We marked each predicted profile with a red dot at the energy corresponding to the observed direct gap and we report a thick horizontal line at $30000 cm^{-1}$ the value of the absorption used in the experiments to extract the value of the direct gap. Our results at lower pressures are in reasonable agreement with this criterion. However at the higher pressure absorption at the energy gap is about 2-3 times higher than $30000 cm^{-1}$. 

{\bf Conclusions.} We have studied the closure of the fundamental gap with pressure of candidate structures of molecular hydrogen in phase III (C2/c-24 and Cmca-12) and phase IV (Pc48) entirely based on Quantum Monte Carlo. 
\CP{For ideal structures our gap values are in excellent agreement with GW prediction\cite{McMinis2015b}.}
Considering quantum nuclei at finite temperature, we observe a strong reduction of the energy gap with respect to the ideal structures at the same pressure ($\sim$ 2eV) caused by ZPM. 
At 200K the fundamental (indirect) gap closes at $\sim$ 370-380GPa for C2/c-24 and at $\sim$ 340GPa for Cmca-12. We observe a reasonable agreement with experimental determinations of indirect gaps from optical absorption. \CP{The direct gap remains open until $\sim$ 450GPa for C2/c-24 and $\sim 500GPa$ for Cmca-12.} 
In this range of pressure the system is a bad metal (or semi-metal) suggesting a scenario that qualitatively supports recent experiments \cite{Dias2016,Eremets2017,Eremets2019,Dias2019}.
In Refs ~\cite{Eremets2017,Eremets2019} no discontinuities in the Raman vibrational spectrum are reported when entering the semi-metallic phase, while in Refs~\cite{Dias2016, Dias2019} new IR vibron peaks are reported in this pressure range and ascribed to a structural phase transition. They have been tentatively assigned to a transition from the C2/c-24 to the Cmca-12 structure\cite{Dias2019}.  
\CP{Our present results, supplemented by free energy calculations  \cite{Pierleoni2020}, do not disprove this hypothesis.}  
\CP{Our predictions for direct gap are in good agreement with the experimental data at T=100K \cite{Loubeyre2002,Loubeyre2019}. However our absorption profiles do not agree as well with the experimental behaviour. This disagreement remains an open question.}
\nocite{MartinReiningCeperley,Perdew1985,Lin2001,Holzmann2009,GrossoBook,WootenBook,Morales2014entropy}

 \begin{acknowledgments}
We thank Paul Loubeyre, Mikhail Eremets, Mario Santoro and Michele Nardone for useful suggestions. D.M.C. was supported by DOE Grant NA DE-NA0001789 and by the Fondation NanoSciences (Grenoble). V.G. and C.P. were supported by the Agence Nationale de la Recherche (ANR) France, under the program ``Accueil de Chercheurs de Haut Niveau 2015'' project: HyLightExtreme. Computer time was provided by the PRACE Project 2016143296, ISCRAB (IsB17\_MMCRHY) computer allocation at CINECA Italy, the high-performance computer resources from Grand Equipement National de Calcul Intensif (GENCI) Allocation 2018-A0030910282, by an allocation of the Blue Waters sustained petascale computing project, supported by the National Science Foundation (Award OCI 07- 25070) and the State of Illinois, and by
the Froggy platform of CIMENT, Grenoble (Rh{\^o}ne-Alpes CPER07-13 CIRA and ANR-10-EQPX-29-01).
\end{acknowledgments}

\bibliographystyle{apsrev4-1}
\bibliography{HPHydrogen}

\begin{thebibliography}{74}%
\makeatletter
\providecommand \@ifxundefined [1]{%
 \@ifx{#1\undefined}
}%
\providecommand \@ifnum [1]{%
 \ifnum #1\expandafter \@firstoftwo
 \else \expandafter \@secondoftwo
 \fi
}%
\providecommand \@ifx [1]{%
 \ifx #1\expandafter \@firstoftwo
 \else \expandafter \@secondoftwo
 \fi
}%
\providecommand \natexlab [1]{#1}%
\providecommand \enquote  [1]{``#1''}%
\providecommand \bibnamefont  [1]{#1}%
\providecommand \bibfnamefont [1]{#1}%
\providecommand \citenamefont [1]{#1}%
\providecommand \href@noop [0]{\@secondoftwo}%
\providecommand \href [0]{\begingroup \@sanitize@url \@href}%
\providecommand \@href[1]{\@@startlink{#1}\@@href}%
\providecommand \@@href[1]{\endgroup#1\@@endlink}%
\providecommand \@sanitize@url [0]{\catcode `\\12\catcode `\$12\catcode
  `\&12\catcode `\#12\catcode `\^12\catcode `\_12\catcode `\%12\relax}%
\providecommand \@@startlink[1]{}%
\providecommand \@@endlink[0]{}%
\providecommand \url  [0]{\begingroup\@sanitize@url \@url }%
\providecommand \@url [1]{\endgroup\@href {#1}{\urlprefix }}%
\providecommand \urlprefix  [0]{URL }%
\providecommand \Eprint [0]{\href }%
\providecommand \doibase [0]{http://dx.doi.org/}%
\providecommand \selectlanguage [0]{\@gobble}%
\providecommand \bibinfo  [0]{\@secondoftwo}%
\providecommand \bibfield  [0]{\@secondoftwo}%
\providecommand \translation [1]{[#1]}%
\providecommand \BibitemOpen [0]{}%
\providecommand \bibitemStop [0]{}%
\providecommand \bibitemNoStop [0]{.\EOS\space}%
\providecommand \EOS [0]{\spacefactor3000\relax}%
\providecommand \BibitemShut  [1]{\csname bibitem#1\endcsname}%
\let\auto@bib@innerbib\@empty
\bibitem [{\citenamefont {Wigner}\ and\ \citenamefont
  {Huntington}(1935)}]{Wigner1935}%
  \BibitemOpen
  \bibfield  {author} {\bibinfo {author} {\bibfnamefont {E.}~\bibnamefont
  {Wigner}}\ and\ \bibinfo {author} {\bibfnamefont {H.~B.}\ \bibnamefont
  {Huntington}},\ }\href {\doibase 10.1063/1.1749590} {\bibfield  {journal}
  {\bibinfo  {journal} {J. Chem. Phys.}\ }\textbf {\bibinfo {volume} {3}},\
  \bibinfo {pages} {764} (\bibinfo {year} {1935})}\BibitemShut {NoStop}%
\bibitem [{\citenamefont {Dias}\ and\ \citenamefont
  {Silvera}(2017)}]{Dias2017}%
  \BibitemOpen
  \bibfield  {author} {\bibinfo {author} {\bibfnamefont {R.~P.}\ \bibnamefont
  {Dias}}\ and\ \bibinfo {author} {\bibfnamefont {I.~F.}\ \bibnamefont
  {Silvera}},\ }\href
  {http://science.sciencemag.org/content/sci/355/6326/715.full.pdf} {\bibfield
  {journal} {\bibinfo  {journal} {Science}\ }\textbf {\bibinfo {volume}
  {355}},\ \bibinfo {pages} {715} (\bibinfo {year} {2017})}\BibitemShut
  {NoStop}%
\bibitem [{\citenamefont {Goncharov}\ and\ \citenamefont
  {Struzhkin}(2017)}]{Goncharov2017}%
  \BibitemOpen
  \bibfield  {author} {\bibinfo {author} {\bibfnamefont {A.~F.}\ \bibnamefont
  {Goncharov}}\ and\ \bibinfo {author} {\bibfnamefont {V.~V.}\ \bibnamefont
  {Struzhkin}},\ }\href {\doibase 10.1126/science.aam9736} {\bibfield
  {journal} {\bibinfo  {journal} {Science}\ }\textbf {\bibinfo {volume} {357}}
  (\bibinfo {year} {2017}),\ 10.1126/science.aam9736}\BibitemShut {NoStop}%
\bibitem [{\citenamefont {Loubeyre}\ \emph {et~al.}(2017)\citenamefont
  {Loubeyre}, \citenamefont {Occelli},\ and\ \citenamefont
  {Dumas}}]{Loubeyre2017}%
  \BibitemOpen
  \bibfield  {author} {\bibinfo {author} {\bibfnamefont {P.}~\bibnamefont
  {Loubeyre}}, \bibinfo {author} {\bibfnamefont {F.}~\bibnamefont {Occelli}}, \
  and\ \bibinfo {author} {\bibfnamefont {P.}~\bibnamefont {Dumas}},\ }\href
  {https://arxiv.org/pdf/1702.07192.pdf} {\  (\bibinfo {year} {2017})},\
  \Eprint {http://arxiv.org/abs/arXiv:1702.07192} {arXiv:1702.07192}
  \BibitemShut {NoStop}%
\bibitem [{\citenamefont {Liu}\ \emph {et~al.}(2017)\citenamefont {Liu},
  \citenamefont {Dalladay-Simpson}, \citenamefont {Howie}, \citenamefont {Li},\
  and\ \citenamefont {Gregoryanz}}]{Liu2017}%
  \BibitemOpen
  \bibfield  {author} {\bibinfo {author} {\bibfnamefont {X.-D.}\ \bibnamefont
  {Liu}}, \bibinfo {author} {\bibfnamefont {P.}~\bibnamefont
  {Dalladay-Simpson}}, \bibinfo {author} {\bibfnamefont {R.~T.}\ \bibnamefont
  {Howie}}, \bibinfo {author} {\bibfnamefont {B.}~\bibnamefont {Li}}, \ and\
  \bibinfo {author} {\bibfnamefont {E.}~\bibnamefont {Gregoryanz}},\ }\href
  {https://arxiv.org/pdf/1704.07601.pdf} {\  (\bibinfo {year} {2017})},\
  \Eprint {http://arxiv.org/abs/arXiv:1704.07601v2} {arXiv:1704.07601v2}
  \BibitemShut {NoStop}%
\bibitem [{\citenamefont {Silvera}\ and\ \citenamefont
  {Dias}(2017)}]{Silvera2017a}%
  \BibitemOpen
  \bibfield  {author} {\bibinfo {author} {\bibfnamefont {I.}~\bibnamefont
  {Silvera}}\ and\ \bibinfo {author} {\bibfnamefont {R.}~\bibnamefont {Dias}},\
  }\href {\doibase 10.1126/science.aan1215} {\bibfield  {journal} {\bibinfo
  {journal} {Science}\ }\textbf {\bibinfo {volume} {357}},\ \bibinfo {pages}
  {eaan1215} (\bibinfo {year} {2017})},\ \Eprint
  {http://arxiv.org/abs/1603.02162} {arXiv:1603.02162} \BibitemShut {NoStop}%
\bibitem [{\citenamefont {Mao}\ and\ \citenamefont {Hemley}(1994)}]{Mao1994}%
  \BibitemOpen
  \bibfield  {author} {\bibinfo {author} {\bibfnamefont {H.~K.}\ \bibnamefont
  {Mao}}\ and\ \bibinfo {author} {\bibfnamefont {R.}~\bibnamefont {Hemley}},\
  }\href@noop {} {\bibfield  {journal} {\bibinfo  {journal} {Rev. Mod. Phys.}\
  }\textbf {\bibinfo {volume} {66}},\ \bibinfo {pages} {671} (\bibinfo {year}
  {1994})}\BibitemShut {NoStop}%
\bibitem [{\citenamefont {McMahon}\ \emph {et~al.}(2012)\citenamefont
  {McMahon}, \citenamefont {Morales}, \citenamefont {Pierleoni},\ and\
  \citenamefont {Ceperley}}]{McMahon2012a}%
  \BibitemOpen
  \bibfield  {author} {\bibinfo {author} {\bibfnamefont {J.~M.}\ \bibnamefont
  {McMahon}}, \bibinfo {author} {\bibfnamefont {M.~A.}\ \bibnamefont
  {Morales}}, \bibinfo {author} {\bibfnamefont {C.}~\bibnamefont {Pierleoni}},
  \ and\ \bibinfo {author} {\bibfnamefont {D.~M.}\ \bibnamefont {Ceperley}},\
  }\href {\doibase 10.1103/RevModPhys.84.1607} {\bibfield  {journal} {\bibinfo
  {journal} {Rev. Mod. Phys.}\ }\textbf {\bibinfo {volume} {84}},\ \bibinfo
  {pages} {1607} (\bibinfo {year} {2012})}\BibitemShut {NoStop}%
\bibitem [{\citenamefont {Nellis}(2006)}]{Nellis2006}%
  \BibitemOpen
  \bibfield  {author} {\bibinfo {author} {\bibfnamefont {W.~J.}\ \bibnamefont
  {Nellis}},\ }\href {http://stacks.iop.org/0034-4885/69/i=5/a=R05} {\bibfield
  {journal} {\bibinfo  {journal} {Reports on Progress in Physics}\ }\textbf
  {\bibinfo {volume} {69}},\ \bibinfo {pages} {1479} (\bibinfo {year}
  {2006})}\BibitemShut {NoStop}%
\bibitem [{\citenamefont {Loubeyre}\ \emph {et~al.}(1996)\citenamefont
  {Loubeyre}, \citenamefont {LeToullec}, \citenamefont {Hausermann},
  \citenamefont {Hanfland}, \citenamefont {Hemley}, \citenamefont {Mao},\ and\
  \citenamefont {Finger}}]{Loubeyre1996}%
  \BibitemOpen
  \bibfield  {author} {\bibinfo {author} {\bibfnamefont {P.}~\bibnamefont
  {Loubeyre}}, \bibinfo {author} {\bibfnamefont {R.}~\bibnamefont {LeToullec}},
  \bibinfo {author} {\bibfnamefont {D.}~\bibnamefont {Hausermann}}, \bibinfo
  {author} {\bibfnamefont {M.}~\bibnamefont {Hanfland}}, \bibinfo {author}
  {\bibfnamefont {R.~J.}\ \bibnamefont {Hemley}}, \bibinfo {author}
  {\bibfnamefont {H.~K.}\ \bibnamefont {Mao}}, \ and\ \bibinfo {author}
  {\bibfnamefont {L.~W.}\ \bibnamefont {Finger}},\ }\href@noop {} {\bibfield
  {journal} {\bibinfo  {journal} {Nature}\ }\textbf {\bibinfo {volume} {383}},\
  \bibinfo {pages} {702} (\bibinfo {year} {1996})}\BibitemShut {NoStop}%
\bibitem [{\citenamefont {Ji}\ \emph {et~al.}(2019)\citenamefont {Ji},
  \citenamefont {Li}, \citenamefont {Liu}, \citenamefont {Smith}, \citenamefont
  {Majumdar}, \citenamefont {Luo}, \citenamefont {Ahuja}, \citenamefont {Shu},
  \citenamefont {Wang}, \citenamefont {Sinogeikin}, \citenamefont {Meng},
  \citenamefont {Prakapenka}, \citenamefont {Greenberg}, \citenamefont {Xu},
  \citenamefont {Huang}, \citenamefont {Yang}, \citenamefont {Shen},
  \citenamefont {Mao},\ and\ \citenamefont {Mao}}]{Ji2019}%
  \BibitemOpen
  \bibfield  {author} {\bibinfo {author} {\bibfnamefont {C.}~\bibnamefont
  {Ji}}, \bibinfo {author} {\bibfnamefont {B.}~\bibnamefont {Li}}, \bibinfo
  {author} {\bibfnamefont {W.}~\bibnamefont {Liu}}, \bibinfo {author}
  {\bibfnamefont {J.~S.}\ \bibnamefont {Smith}}, \bibinfo {author}
  {\bibfnamefont {A.}~\bibnamefont {Majumdar}}, \bibinfo {author}
  {\bibfnamefont {W.}~\bibnamefont {Luo}}, \bibinfo {author} {\bibfnamefont
  {R.}~\bibnamefont {Ahuja}}, \bibinfo {author} {\bibfnamefont
  {J.}~\bibnamefont {Shu}}, \bibinfo {author} {\bibfnamefont {J.}~\bibnamefont
  {Wang}}, \bibinfo {author} {\bibfnamefont {S.}~\bibnamefont {Sinogeikin}},
  \bibinfo {author} {\bibfnamefont {Y.}~\bibnamefont {Meng}}, \bibinfo {author}
  {\bibfnamefont {V.~B.}\ \bibnamefont {Prakapenka}}, \bibinfo {author}
  {\bibfnamefont {E.}~\bibnamefont {Greenberg}}, \bibinfo {author}
  {\bibfnamefont {R.}~\bibnamefont {Xu}}, \bibinfo {author} {\bibfnamefont
  {X.}~\bibnamefont {Huang}}, \bibinfo {author} {\bibfnamefont
  {W.}~\bibnamefont {Yang}}, \bibinfo {author} {\bibfnamefont {G.}~\bibnamefont
  {Shen}}, \bibinfo {author} {\bibfnamefont {W.~L.}\ \bibnamefont {Mao}}, \
  and\ \bibinfo {author} {\bibfnamefont {H.-K.}\ \bibnamefont {Mao}},\ }\href
  {\doibase 10.1038/s41586-019-1565-9} {\bibfield  {journal} {\bibinfo
  {journal} {Nature}\ }\textbf {\bibinfo {volume} {573}},\ \bibinfo {pages}
  {558} (\bibinfo {year} {2019})}\BibitemShut {NoStop}%
\bibitem [{\citenamefont {Dubrovinsky}\ \emph {et~al.}(2019)\citenamefont
  {Dubrovinsky}, \citenamefont {Dubrovinskaia},\ and\ \citenamefont
  {Katsnelson}}]{Dubrovinsky2019}%
  \BibitemOpen
  \bibfield  {author} {\bibinfo {author} {\bibfnamefont {L.}~\bibnamefont
  {Dubrovinsky}}, \bibinfo {author} {\bibfnamefont {N.}~\bibnamefont
  {Dubrovinskaia}}, \ and\ \bibinfo {author} {\bibfnamefont {M.~I.}\
  \bibnamefont {Katsnelson}},\ }\href@noop {} {\bibfield  {journal} {\bibinfo
  {journal} {arxiv:1910.10772}\ } (\bibinfo {year} {2019})}\BibitemShut
  {NoStop}%
\bibitem [{\citenamefont {Silvera}\ and\ \citenamefont
  {Dias}(2018)}]{Silvera2018}%
  \BibitemOpen
  \bibfield  {author} {\bibinfo {author} {\bibfnamefont {I.}~\bibnamefont
  {Silvera}}\ and\ \bibinfo {author} {\bibfnamefont {R.~P.}\ \bibnamefont
  {Dias}},\ }\href {\doibase 10.1088/1361-648X/aac401} {\bibfield  {journal}
  {\bibinfo  {journal} {J. Phys.: Condens. Matter}\ }\textbf {\bibinfo {volume}
  {30}},\ \bibinfo {pages} {254003} (\bibinfo {year} {2018})}\BibitemShut
  {NoStop}%
\bibitem [{\citenamefont {Eremets}\ and\ \citenamefont
  {Troyan}(2011)}]{Eremets2011}%
  \BibitemOpen
  \bibfield  {author} {\bibinfo {author} {\bibfnamefont {M.~I.}\ \bibnamefont
  {Eremets}}\ and\ \bibinfo {author} {\bibfnamefont {I.~A.}\ \bibnamefont
  {Troyan}},\ }\href {\doibase 10.1038/nmat3175} {\bibfield  {journal}
  {\bibinfo  {journal} {Nature materials}\ }\textbf {\bibinfo {volume} {10}},\
  \bibinfo {pages} {927} (\bibinfo {year} {2011})}\BibitemShut {NoStop}%
\bibitem [{\citenamefont {Nellis}\ \emph {et~al.}(2012)\citenamefont {Nellis},
  \citenamefont {Ruoff},\ and\ \citenamefont {Silvera}}]{Nellis2012}%
  \BibitemOpen
  \bibfield  {author} {\bibinfo {author} {\bibfnamefont {W.~J.}\ \bibnamefont
  {Nellis}}, \bibinfo {author} {\bibfnamefont {A.~L.}\ \bibnamefont {Ruoff}}, \
  and\ \bibinfo {author} {\bibfnamefont {I.~F.}\ \bibnamefont {Silvera}},\
  }\href {http://arxiv.org/abs/1201.0407} {\  (\bibinfo {year} {2012})},\
  \Eprint {http://arxiv.org/abs/1201.0407} {arXiv:1201.0407} \BibitemShut
  {NoStop}%
\bibitem [{\citenamefont {Goncharov}\ and\ \citenamefont
  {Struzhkin}(2012)}]{Goncharov2012}%
  \BibitemOpen
  \bibfield  {author} {\bibinfo {author} {\bibfnamefont {A.~F.}\ \bibnamefont
  {Goncharov}}\ and\ \bibinfo {author} {\bibfnamefont {V.~V.}\ \bibnamefont
  {Struzhkin}},\ }\href
  {papers://68dd8153-82a1-4a0e-9ae0-80ae65a14bf7/Paper/p2706} {\bibfield
  {journal} {\bibinfo  {journal} {Nature Materials}\ }\textbf {\bibinfo
  {volume} {10}},\ \bibinfo {pages} {927} (\bibinfo {year} {2012})}\BibitemShut
  {NoStop}%
\bibitem [{\citenamefont {Goncharov}\ \emph {et~al.}(2013)\citenamefont
  {Goncharov}, \citenamefont {Tse}, \citenamefont {Wang}, \citenamefont {Yang},
  \citenamefont {Struzhkin}, \citenamefont {Howie},\ and\ \citenamefont
  {Gregoryanz}}]{Goncharov2013}%
  \BibitemOpen
  \bibfield  {author} {\bibinfo {author} {\bibfnamefont {A.~F.}\ \bibnamefont
  {Goncharov}}, \bibinfo {author} {\bibfnamefont {J.~S.}\ \bibnamefont {Tse}},
  \bibinfo {author} {\bibfnamefont {H.}~\bibnamefont {Wang}}, \bibinfo {author}
  {\bibfnamefont {J.}~\bibnamefont {Yang}}, \bibinfo {author} {\bibfnamefont
  {V.~V.}\ \bibnamefont {Struzhkin}}, \bibinfo {author} {\bibfnamefont {R.~T.}\
  \bibnamefont {Howie}}, \ and\ \bibinfo {author} {\bibfnamefont
  {E.}~\bibnamefont {Gregoryanz}},\ }\href {\doibase
  10.1103/PhysRevB.87.024101} {\bibfield  {journal} {\bibinfo  {journal} {Phys.
  Rev. B}\ }\textbf {\bibinfo {volume} {87}},\ \bibinfo {pages} {024101}
  (\bibinfo {year} {2013})}\BibitemShut {NoStop}%
\bibitem [{\citenamefont {Eremets}\ \emph {et~al.}(2016)\citenamefont
  {Eremets}, \citenamefont {Troyan},\ and\ \citenamefont
  {Drozdov}}]{Eremets2016}%
  \BibitemOpen
  \bibfield  {author} {\bibinfo {author} {\bibfnamefont {M.~I.}\ \bibnamefont
  {Eremets}}, \bibinfo {author} {\bibfnamefont {I.~A.}\ \bibnamefont {Troyan}},
  \ and\ \bibinfo {author} {\bibfnamefont {A.~P.}\ \bibnamefont {Drozdov}},\
  }\href {http://arxiv.org/abs/1601.04479} {\  (\bibinfo {year} {2016})},\
  \Eprint {http://arxiv.org/abs/1601.04479} {arXiv:1601.04479} \BibitemShut
  {NoStop}%
\bibitem [{\citenamefont {Eremets}\ \emph {et~al.}(2017)\citenamefont
  {Eremets}, \citenamefont {Drozdov}, \citenamefont {Kong},\ and\ \citenamefont
  {Wang}}]{Eremets2017}%
  \BibitemOpen
  \bibfield  {author} {\bibinfo {author} {\bibfnamefont {M.}~\bibnamefont
  {Eremets}}, \bibinfo {author} {\bibfnamefont {A.~P.}\ \bibnamefont
  {Drozdov}}, \bibinfo {author} {\bibfnamefont {P.}~\bibnamefont {Kong}}, \
  and\ \bibinfo {author} {\bibfnamefont {H.}~\bibnamefont {Wang}},\ }\href
  {https://arxiv.org/abs/1708.05217} {\  (\bibinfo {year} {2017})},\ \Eprint
  {http://arxiv.org/abs/https://arxiv.org/abs/1708.05217}
  {https://arxiv.org/abs/1708.05217} \BibitemShut {NoStop}%
\bibitem [{\citenamefont {Dias}\ \emph {et~al.}(2016)\citenamefont {Dias},
  \citenamefont {Noked},\ and\ \citenamefont {Silvera}}]{Dias2016}%
  \BibitemOpen
  \bibfield  {author} {\bibinfo {author} {\bibfnamefont {R.~P.}\ \bibnamefont
  {Dias}}, \bibinfo {author} {\bibfnamefont {O.}~\bibnamefont {Noked}}, \ and\
  \bibinfo {author} {\bibfnamefont {I.~F.}\ \bibnamefont {Silvera}},\
  }\href@noop {} {\  (\bibinfo {year} {2016})},\ \Eprint
  {http://arxiv.org/abs/1603.02162} {arXiv:1603.02162} \BibitemShut {NoStop}%
\bibitem [{\citenamefont {Loubeyre}\ \emph {et~al.}(2020)\citenamefont
  {Loubeyre}, \citenamefont {Occelli},\ and\ \citenamefont
  {Dumas}}]{Loubeyre2019}%
  \BibitemOpen
  \bibfield  {author} {\bibinfo {author} {\bibfnamefont {P.}~\bibnamefont
  {Loubeyre}}, \bibinfo {author} {\bibfnamefont {F.}~\bibnamefont {Occelli}}, \
  and\ \bibinfo {author} {\bibfnamefont {P.}~\bibnamefont {Dumas}},\ }\href
  {\doibase 10.1038/s41586-019-1927-3} {\bibfield  {journal} {\bibinfo
  {journal} {Nature}\ }\textbf {\bibinfo {volume} {577}},\ \bibinfo {pages}
  {631} (\bibinfo {year} {2020})}\BibitemShut {NoStop}%
\bibitem [{\citenamefont {Dias}\ \emph {et~al.}(2019)\citenamefont {Dias},
  \citenamefont {Noked},\ and\ \citenamefont {Silvera}}]{Dias2019}%
  \BibitemOpen
  \bibfield  {author} {\bibinfo {author} {\bibfnamefont {R.~P.}\ \bibnamefont
  {Dias}}, \bibinfo {author} {\bibfnamefont {O.}~\bibnamefont {Noked}}, \ and\
  \bibinfo {author} {\bibfnamefont {I.~F.}\ \bibnamefont {Silvera}},\ }\href
  {\doibase 10.1103/PhysRevB.100.184112} {\bibfield  {journal} {\bibinfo
  {journal} {Physical Review B}\ }\textbf {\bibinfo {volume} {100}},\ \bibinfo
  {pages} {184112} (\bibinfo {year} {2019})}\BibitemShut {NoStop}%
\bibitem [{\citenamefont {Howie}\ \emph
  {et~al.}(2012{\natexlab{a}})\citenamefont {Howie}, \citenamefont {Scheler},
  \citenamefont {Guillaume},\ and\ \citenamefont {Gregoryanz}}]{Howie2012}%
  \BibitemOpen
  \bibfield  {author} {\bibinfo {author} {\bibfnamefont {R.~T.}\ \bibnamefont
  {Howie}}, \bibinfo {author} {\bibfnamefont {T.}~\bibnamefont {Scheler}},
  \bibinfo {author} {\bibfnamefont {C.~L.}\ \bibnamefont {Guillaume}}, \ and\
  \bibinfo {author} {\bibfnamefont {E.}~\bibnamefont {Gregoryanz}},\ }\href
  {papers://68dd8153-82a1-4a0e-9ae0-80ae65a14bf7/Paper/p3005} {\bibfield
  {journal} {\bibinfo  {journal} {Physical Review B}\ }\textbf {\bibinfo
  {volume} {86}},\ \bibinfo {pages} {214104} (\bibinfo {year}
  {2012}{\natexlab{a}})}\BibitemShut {NoStop}%
\bibitem [{\citenamefont {Howie}\ \emph
  {et~al.}(2012{\natexlab{b}})\citenamefont {Howie}, \citenamefont {Guillaume},
  \citenamefont {Scheler}, \citenamefont {Goncharov},\ and\ \citenamefont
  {Gregoryanz}}]{Howie2012b}%
  \BibitemOpen
  \bibfield  {author} {\bibinfo {author} {\bibfnamefont {R.~T.}\ \bibnamefont
  {Howie}}, \bibinfo {author} {\bibfnamefont {C.~L.}\ \bibnamefont
  {Guillaume}}, \bibinfo {author} {\bibfnamefont {T.}~\bibnamefont {Scheler}},
  \bibinfo {author} {\bibfnamefont {A.~F.}\ \bibnamefont {Goncharov}}, \ and\
  \bibinfo {author} {\bibfnamefont {E.}~\bibnamefont {Gregoryanz}},\ }\href
  {\doibase 10.1103/PhysRevLett.108.125501} {\bibfield  {journal} {\bibinfo
  {journal} {Phys. Rev. Letts.}\ }\textbf {\bibinfo {volume} {108}},\ \bibinfo
  {pages} {125501} (\bibinfo {year} {2012}{\natexlab{b}})}\BibitemShut
  {NoStop}%
\bibitem [{\citenamefont {Zha}\ \emph {et~al.}(2012)\citenamefont {Zha},
  \citenamefont {Liu},\ and\ \citenamefont {Hemley}}]{Zha2012}%
  \BibitemOpen
  \bibfield  {author} {\bibinfo {author} {\bibfnamefont {C.~S.}\ \bibnamefont
  {Zha}}, \bibinfo {author} {\bibfnamefont {Z.}~\bibnamefont {Liu}}, \ and\
  \bibinfo {author} {\bibfnamefont {R.~J.}\ \bibnamefont {Hemley}},\ }\href
  {\doibase 10.1103/PhysRevLett.108.146402} {\bibfield  {journal} {\bibinfo
  {journal} {Phys. Rev. Letts.}\ }\textbf {\bibinfo {volume} {108}},\ \bibinfo
  {pages} {146402} (\bibinfo {year} {2012})}\BibitemShut {NoStop}%
\bibitem [{\citenamefont {Loubeyre}\ \emph {et~al.}(2013)\citenamefont
  {Loubeyre}, \citenamefont {Occelli},\ and\ \citenamefont
  {Dumas}}]{Loubeyre2013}%
  \BibitemOpen
  \bibfield  {author} {\bibinfo {author} {\bibfnamefont {P.}~\bibnamefont
  {Loubeyre}}, \bibinfo {author} {\bibfnamefont {F.}~\bibnamefont {Occelli}}, \
  and\ \bibinfo {author} {\bibfnamefont {P.}~\bibnamefont {Dumas}},\ }\href
  {\doibase 10.1103/PhysRevB.87.134101} {\bibfield  {journal} {\bibinfo
  {journal} {Phys. Rev. B}\ }\textbf {\bibinfo {volume} {87}},\ \bibinfo
  {pages} {134101} (\bibinfo {year} {2013})}\BibitemShut {NoStop}%
\bibitem [{\citenamefont {Howie}\ \emph {et~al.}(2015)\citenamefont {Howie},
  \citenamefont {Dalladay-Simpson},\ and\ \citenamefont
  {Gregoryanz}}]{Howie2015}%
  \BibitemOpen
  \bibfield  {author} {\bibinfo {author} {\bibfnamefont {R.~T.}\ \bibnamefont
  {Howie}}, \bibinfo {author} {\bibfnamefont {P.}~\bibnamefont
  {Dalladay-Simpson}}, \ and\ \bibinfo {author} {\bibfnamefont
  {E.}~\bibnamefont {Gregoryanz}},\ }\href {\doibase 10.1038/nmat4213}
  {\bibfield  {journal} {\bibinfo  {journal} {Nature Materials}\ }\textbf
  {\bibinfo {volume} {14}},\ \bibinfo {pages} {1} (\bibinfo {year}
  {2015})}\BibitemShut {NoStop}%
\bibitem [{\citenamefont {Eremets}\ \emph {et~al.}(2019)\citenamefont
  {Eremets}, \citenamefont {Drozdov}, \citenamefont {Kong},\ and\ \citenamefont
  {Wang}}]{Eremets2019}%
  \BibitemOpen
  \bibfield  {author} {\bibinfo {author} {\bibfnamefont {M.~I.}\ \bibnamefont
  {Eremets}}, \bibinfo {author} {\bibfnamefont {A.~P.}\ \bibnamefont
  {Drozdov}}, \bibinfo {author} {\bibfnamefont {P.~P.}\ \bibnamefont {Kong}}, \
  and\ \bibinfo {author} {\bibfnamefont {H.}~\bibnamefont {Wang}},\ }\href
  {\doibase 10.1038/s41567-019-0646-x} {\bibfield  {journal} {\bibinfo
  {journal} {Nat. Phys.}\ } (\bibinfo {year} {2019}),\
  10.1038/s41567-019-0646-x}\BibitemShut {NoStop}%
\bibitem [{\citenamefont {Eremets}\ and\ \citenamefont
  {Drozdov}(2016)}]{Eremets2016a}%
  \BibitemOpen
  \bibfield  {author} {\bibinfo {author} {\bibfnamefont {M.~I.}\ \bibnamefont
  {Eremets}}\ and\ \bibinfo {author} {\bibfnamefont {A.~P.}\ \bibnamefont
  {Drozdov}},\ }\href {http://arxiv.org/abs/1601.04479} {\  (\bibinfo {year}
  {2016})},\ \Eprint {http://arxiv.org/abs/1601.04479} {arXiv:1601.04479}
  \BibitemShut {NoStop}%
\bibitem [{\citenamefont {Silvera}\ and\ \citenamefont
  {Dias}(2019)}]{Silvera2019}%
  \BibitemOpen
  \bibfield  {author} {\bibinfo {author} {\bibfnamefont {I.~F.}\ \bibnamefont
  {Silvera}}\ and\ \bibinfo {author} {\bibfnamefont {R.~P.}\ \bibnamefont
  {Dias}},\ }\href {http://arxiv.org/abs/1906.05634} {\bibfield  {journal}
  {\bibinfo  {journal} {arXiv}\ } (\bibinfo {year} {2019})},\ \Eprint
  {http://arxiv.org/abs/1906.05634} {arXiv:1906.05634} \BibitemShut {NoStop}%
\bibitem [{\citenamefont {Pickard}\ and\ \citenamefont
  {Needs}(2007)}]{Pickard2007}%
  \BibitemOpen
  \bibfield  {author} {\bibinfo {author} {\bibfnamefont {C.~J.}\ \bibnamefont
  {Pickard}}\ and\ \bibinfo {author} {\bibfnamefont {R.~J.}\ \bibnamefont
  {Needs}},\ }\href {\doibase 10.1038/nphys625} {\bibfield  {journal} {\bibinfo
   {journal} {Nature Physics}\ }\textbf {\bibinfo {volume} {3}},\ \bibinfo
  {pages} {473} (\bibinfo {year} {2007})}\BibitemShut {NoStop}%
\bibitem [{\citenamefont {Pickard}\ \emph
  {et~al.}(2012{\natexlab{a}})\citenamefont {Pickard}, \citenamefont
  {Martinez-Canales},\ and\ \citenamefont {Needs}}]{Pickard2012}%
  \BibitemOpen
  \bibfield  {author} {\bibinfo {author} {\bibfnamefont {C.~J.}\ \bibnamefont
  {Pickard}}, \bibinfo {author} {\bibfnamefont {M.}~\bibnamefont
  {Martinez-Canales}}, \ and\ \bibinfo {author} {\bibfnamefont {R.~J.}\
  \bibnamefont {Needs}},\ }\href {\doibase 10.1103/PhysRevB.85.214114}
  {\bibfield  {journal} {\bibinfo  {journal} {Physical Review B}\ }\textbf
  {\bibinfo {volume} {85}},\ \bibinfo {pages} {214114} (\bibinfo {year}
  {2012}{\natexlab{a}})}\BibitemShut {NoStop}%
\bibitem [{\citenamefont {Yang}\ \emph {et~al.}(2020)\citenamefont {Yang},
  \citenamefont {Gorelov}, \citenamefont {Pierleoni}, \citenamefont
  {Ceperley},\ and\ \citenamefont {Holzmann}}]{Yang2019}%
  \BibitemOpen
  \bibfield  {author} {\bibinfo {author} {\bibfnamefont {Y.}~\bibnamefont
  {Yang}}, \bibinfo {author} {\bibfnamefont {V.}~\bibnamefont {Gorelov}},
  \bibinfo {author} {\bibfnamefont {C.}~\bibnamefont {Pierleoni}}, \bibinfo
  {author} {\bibfnamefont {D.~M.}\ \bibnamefont {Ceperley}}, \ and\ \bibinfo
  {author} {\bibfnamefont {M.}~\bibnamefont {Holzmann}},\ }\href@noop {}
  {\bibfield  {journal} {\bibinfo  {journal} {Phys. Rev. B}\ }\textbf {\bibinfo
  {volume} {101}},\ \bibinfo {pages} {085115} (\bibinfo {year}
  {2020})}\BibitemShut {NoStop}%
\bibitem [{\citenamefont {see Supplementary~Material}()}]{SupMat}%
  \BibitemOpen
  \bibfield  {author} {\bibinfo {author} {\bibnamefont {see
  Supplementary~Material}},\ }\href@noop {} {\bibinfo  {journal} {[url] which
  includes Refs. [67-74]}\ }\BibitemShut {NoStop}%
\bibitem [{\citenamefont {Pickard}\ \emph
  {et~al.}(2012{\natexlab{b}})\citenamefont {Pickard}, \citenamefont
  {Martinez-Canales},\ and\ \citenamefont {Needs}}]{Pickard2012erratum}%
  \BibitemOpen
\bibfield  {journal} {  }\bibfield  {author} {\bibinfo {author} {\bibfnamefont
  {C.}~\bibnamefont {Pickard}}, \bibinfo {author} {\bibfnamefont
  {M.}~\bibnamefont {Martinez-Canales}}, \ and\ \bibinfo {author}
  {\bibfnamefont {R.}~\bibnamefont {Needs}},\ }\href {\doibase
  10.1103/PhysRevB.86.059902} {\bibfield  {journal} {\bibinfo  {journal} {Phys.
  Rev. B}\ }\textbf {\bibinfo {volume} {86}},\ \bibinfo {pages} {214114}
  (\bibinfo {year} {2012}{\natexlab{b}})}\BibitemShut {NoStop}%
\bibitem [{\citenamefont {Monserrat}\ \emph {et~al.}(2018)\citenamefont
  {Monserrat}, \citenamefont {Drummond}, \citenamefont {Dalladay-simpson},
  \citenamefont {Howie}, \citenamefont {{Lopez Rios}}, \citenamefont
  {Gregoryanz}, \citenamefont {Pickard},\ and\ \citenamefont
  {Needs}}]{Monserrat2018}%
  \BibitemOpen
  \bibfield  {author} {\bibinfo {author} {\bibfnamefont {B.}~\bibnamefont
  {Monserrat}}, \bibinfo {author} {\bibfnamefont {N.~D.}\ \bibnamefont
  {Drummond}}, \bibinfo {author} {\bibfnamefont {P.}~\bibnamefont
  {Dalladay-simpson}}, \bibinfo {author} {\bibfnamefont {R.~T.}\ \bibnamefont
  {Howie}}, \bibinfo {author} {\bibfnamefont {P.}~\bibnamefont {{Lopez Rios}}},
  \bibinfo {author} {\bibfnamefont {E.}~\bibnamefont {Gregoryanz}}, \bibinfo
  {author} {\bibfnamefont {C.~J.}\ \bibnamefont {Pickard}}, \ and\ \bibinfo
  {author} {\bibfnamefont {R.~J.}\ \bibnamefont {Needs}},\ }\href@noop {}
  {\bibfield  {journal} {\bibinfo  {journal} {Phys Rev Letts}\ }\textbf
  {\bibinfo {volume} {120}},\ \bibinfo {pages} {255701} (\bibinfo {year}
  {2018})}\BibitemShut {NoStop}%
\bibitem [{\citenamefont {McMinis}\ \emph {et~al.}(2015)\citenamefont
  {McMinis}, \citenamefont {Clay}, \citenamefont {Lee},\ and\ \citenamefont
  {Morales}}]{McMinis2015b}%
  \BibitemOpen
  \bibfield  {author} {\bibinfo {author} {\bibfnamefont {J.}~\bibnamefont
  {McMinis}}, \bibinfo {author} {\bibfnamefont {R.~C.}\ \bibnamefont {Clay}},
  \bibinfo {author} {\bibfnamefont {D.}~\bibnamefont {Lee}}, \ and\ \bibinfo
  {author} {\bibfnamefont {M.~A.}\ \bibnamefont {Morales}},\ }\href {\doibase
  10.1103/PhysRevLett.114.105305} {\bibfield  {journal} {\bibinfo  {journal}
  {Phys. Rev. Letts.}\ }\textbf {\bibinfo {volume} {114}},\ \bibinfo {pages}
  {105305} (\bibinfo {year} {2015})}\BibitemShut {NoStop}%
\bibitem [{\citenamefont {Azadi}\ \emph {et~al.}(2014)\citenamefont {Azadi},
  \citenamefont {Monserrat}, \citenamefont {Foulkes},\ and\ \citenamefont
  {Needs}}]{Azadi2014}%
  \BibitemOpen
  \bibfield  {author} {\bibinfo {author} {\bibfnamefont {S.}~\bibnamefont
  {Azadi}}, \bibinfo {author} {\bibfnamefont {B.}~\bibnamefont {Monserrat}},
  \bibinfo {author} {\bibfnamefont {W.~M.~C.}\ \bibnamefont {Foulkes}}, \ and\
  \bibinfo {author} {\bibfnamefont {R.~J.}\ \bibnamefont {Needs}},\ }\href
  {\doibase 10.1103/PhysRevLett.112.165501} {\bibfield  {journal} {\bibinfo
  {journal} {Phys. Rev. Lett.}\ }\textbf {\bibinfo {volume} {112}},\ \bibinfo
  {pages} {165501} (\bibinfo {year} {2014})}\BibitemShut {NoStop}%
\bibitem [{Note1()}]{Note1}%
  \BibitemOpen
  \bibinfo {note} {We have checked that the stress tensor in the constant
  volume CEIMC run remains diagonal with same diagonal elements within our
  statistical noise.}\BibitemShut {Stop}%
\bibitem [{\citenamefont {Pierleoni}\ \emph {et~al.}(2016)\citenamefont
  {Pierleoni}, \citenamefont {Morales}, \citenamefont {Rillo}, \citenamefont
  {Holzmann},\ and\ \citenamefont {Ceperley}}]{Pierleoni2016}%
  \BibitemOpen
  \bibfield  {author} {\bibinfo {author} {\bibfnamefont {C.}~\bibnamefont
  {Pierleoni}}, \bibinfo {author} {\bibfnamefont {M.~A.}\ \bibnamefont
  {Morales}}, \bibinfo {author} {\bibfnamefont {G.}~\bibnamefont {Rillo}},
  \bibinfo {author} {\bibfnamefont {M.}~\bibnamefont {Holzmann}}, \ and\
  \bibinfo {author} {\bibfnamefont {D.~M.}\ \bibnamefont {Ceperley}},\ }\href
  {\doibase 10.1073/pnas.1603853113} {\bibfield  {journal} {\bibinfo  {journal}
  {Proc. Natl. Acad. Sci.}\ }\textbf {\bibinfo {volume} {113}},\ \bibinfo
  {pages} {4954} (\bibinfo {year} {2016})}\BibitemShut {NoStop}%
\bibitem [{\citenamefont {Rillo}\ \emph {et~al.}(2018)\citenamefont {Rillo},
  \citenamefont {Morales}, \citenamefont {Ceperley},\ and\ \citenamefont
  {Pierleoni}}]{Rillo2018}%
  \BibitemOpen
  \bibfield  {author} {\bibinfo {author} {\bibfnamefont {G.}~\bibnamefont
  {Rillo}}, \bibinfo {author} {\bibfnamefont {M.~A.}\ \bibnamefont {Morales}},
  \bibinfo {author} {\bibfnamefont {D.~M.}\ \bibnamefont {Ceperley}}, \ and\
  \bibinfo {author} {\bibfnamefont {C.}~\bibnamefont {Pierleoni}},\ }\href
  {http://arxiv.org/abs/1708.07344} {\bibfield  {journal} {\bibinfo  {journal}
  {J. Chem Phys.}\ }\textbf {\bibinfo {volume} {148}},\ \bibinfo {pages}
  {102314} (\bibinfo {year} {2018})}\BibitemShut {NoStop}%
\bibitem [{\citenamefont {Azadi}\ \emph {et~al.}(2017)\citenamefont {Azadi},
  \citenamefont {Drummond},\ and\ \citenamefont {Foulkes}}]{Azadi2017}%
  \BibitemOpen
  \bibfield  {author} {\bibinfo {author} {\bibfnamefont {S.}~\bibnamefont
  {Azadi}}, \bibinfo {author} {\bibfnamefont {N.~D.}\ \bibnamefont {Drummond}},
  \ and\ \bibinfo {author} {\bibfnamefont {W.~M.~C.}\ \bibnamefont {Foulkes}},\
  }\href@noop {} {\bibfield  {journal} {\bibinfo  {journal} {Phys. Rev. B}\
  }\textbf {\bibinfo {volume} {95}},\ \bibinfo {pages} {035142} (\bibinfo
  {year} {2017})}\BibitemShut {NoStop}%
\bibitem [{\citenamefont {Leb{\`{e}}gue}\ \emph {et~al.}(2012)\citenamefont
  {Leb{\`{e}}gue}, \citenamefont {Araujo}, \citenamefont {Kim}, \citenamefont
  {Ramzan}, \citenamefont {Mao},\ and\ \citenamefont {Ahuja}}]{Lebegue2012}%
  \BibitemOpen
  \bibfield  {author} {\bibinfo {author} {\bibfnamefont {S.}~\bibnamefont
  {Leb{\`{e}}gue}}, \bibinfo {author} {\bibfnamefont {C.}~\bibnamefont
  {Araujo}}, \bibinfo {author} {\bibfnamefont {D.}~\bibnamefont {Kim}},
  \bibinfo {author} {\bibfnamefont {M.}~\bibnamefont {Ramzan}}, \bibinfo
  {author} {\bibfnamefont {H.}~\bibnamefont {Mao}}, \ and\ \bibinfo {author}
  {\bibfnamefont {R.}~\bibnamefont {Ahuja}},\ }\href
  {papers://68dd8153-82a1-4a0e-9ae0-80ae65a14bf7/Paper/p2763} {\bibfield
  {journal} {\bibinfo  {journal} {Proceedings of the National Academy of
  Sciences}\ }\textbf {\bibinfo {volume} {109}},\ \bibinfo {pages} {9766}
  (\bibinfo {year} {2012})}\BibitemShut {NoStop}%
\bibitem [{\citenamefont {Wai-Leung~Yim}\ and\ \citenamefont
  {Tse}(2017)}]{Yim}%
  \BibitemOpen
  \bibfield  {author} {\bibinfo {author} {\bibfnamefont {Y.~L. R. J.~H.}\
  \bibnamefont {Wai-Leung~Yim}, \bibfnamefont {Hongliang~Shi}}\ and\ \bibinfo
  {author} {\bibfnamefont {J.~S.}\ \bibnamefont {Tse}},\ }in\ \href@noop {}
  {\emph {\bibinfo {booktitle} {Correlations in Condensed Matter under Extreme
  Conditions}}},\ \bibinfo {editor} {edited by\ \bibinfo {editor}
  {\bibfnamefont {G.~A.}\ \bibnamefont {Editors}}\ and\ \bibinfo {editor}
  {\bibfnamefont {A.~L.}\ \bibnamefont {Magna}}}\ (\bibinfo  {publisher}
  {Springer International Publishing},\ \bibinfo {address} {AG},\ \bibinfo
  {year} {2017})\ Chap.~\bibinfo {chapter} {9}, pp.\ \bibinfo {pages}
  {107--126}\BibitemShut {NoStop}%
\bibitem [{\citenamefont {Dvorak}\ \emph {et~al.}(2014)\citenamefont {Dvorak},
  \citenamefont {Chen},\ and\ \citenamefont {Wu}}]{Dvorak2014}%
  \BibitemOpen
  \bibfield  {author} {\bibinfo {author} {\bibfnamefont {M.}~\bibnamefont
  {Dvorak}}, \bibinfo {author} {\bibfnamefont {X.-J.}\ \bibnamefont {Chen}}, \
  and\ \bibinfo {author} {\bibfnamefont {Z.}~\bibnamefont {Wu}},\ }\href
  {\doibase 10.1103/PhysRevB.90.035103} {\bibfield  {journal} {\bibinfo
  {journal} {Physical Review B}\ }\textbf {\bibinfo {volume} {90}},\ \bibinfo
  {pages} {035103} (\bibinfo {year} {2014})}\BibitemShut {NoStop}%
\bibitem [{Note2()}]{Note2}%
  \BibitemOpen
  \bibinfo {note} {The observed small difference, in particular at the higher
  pressure, is probably due to the different XC approximation used for geometry
  optimization, vdW-DF in our case, BLYP in Ref. \cite {Azadi2017} and
  different size extrapolation.}\BibitemShut {Stop}%
\bibitem [{\citenamefont {Morales}\ \emph {et~al.}(2013)\citenamefont
  {Morales}, \citenamefont {McMahon}, \citenamefont {Pierleoni},\ and\
  \citenamefont {Ceperley}}]{Morales2013}%
  \BibitemOpen
  \bibfield  {author} {\bibinfo {author} {\bibfnamefont {M.~A.}\ \bibnamefont
  {Morales}}, \bibinfo {author} {\bibfnamefont {J.~M.}\ \bibnamefont
  {McMahon}}, \bibinfo {author} {\bibfnamefont {C.}~\bibnamefont {Pierleoni}},
  \ and\ \bibinfo {author} {\bibfnamefont {D.~M.}\ \bibnamefont {Ceperley}},\
  }\href {\doibase 10.1103/PhysRevB.87.184107} {\bibfield  {journal} {\bibinfo
  {journal} {Phys. Rev. B}\ }\textbf {\bibinfo {volume} {87}},\ \bibinfo
  {pages} {184107} (\bibinfo {year} {2013})}\BibitemShut {NoStop}%
\bibitem [{\citenamefont {Clay}\ \emph {et~al.}(2014)\citenamefont {Clay},
  \citenamefont {McMinis}, \citenamefont {McMahon}, \citenamefont {Pierleoni},
  \citenamefont {Ceperley},\ and\ \citenamefont {Morales}}]{Clay2014}%
  \BibitemOpen
  \bibfield  {author} {\bibinfo {author} {\bibfnamefont {R.~C.}\ \bibnamefont
  {Clay}}, \bibinfo {author} {\bibfnamefont {J.}~\bibnamefont {McMinis}},
  \bibinfo {author} {\bibfnamefont {J.~M.}\ \bibnamefont {McMahon}}, \bibinfo
  {author} {\bibfnamefont {C.}~\bibnamefont {Pierleoni}}, \bibinfo {author}
  {\bibfnamefont {D.~M.}\ \bibnamefont {Ceperley}}, \ and\ \bibinfo {author}
  {\bibfnamefont {M.~A.}\ \bibnamefont {Morales}},\ }\href {\doibase
  10.1103/PhysRevB.89.184106} {\bibfield  {journal} {\bibinfo  {journal} {Phys.
  Rev. B}\ }\textbf {\bibinfo {volume} {89}},\ \bibinfo {pages} {184106}
  (\bibinfo {year} {2014})}\BibitemShut {NoStop}%
\bibitem [{\citenamefont {Heyd}\ \emph {et~al.}(2005)\citenamefont {Heyd},
  \citenamefont {Peralta}, \citenamefont {Scuseria},\ and\ \citenamefont
  {Martin}}]{Heyd2005}%
  \BibitemOpen
  \bibfield  {author} {\bibinfo {author} {\bibfnamefont {J.}~\bibnamefont
  {Heyd}}, \bibinfo {author} {\bibfnamefont {J.~E.}\ \bibnamefont {Peralta}},
  \bibinfo {author} {\bibfnamefont {G.~E.}\ \bibnamefont {Scuseria}}, \ and\
  \bibinfo {author} {\bibfnamefont {R.~L.}\ \bibnamefont {Martin}},\ }\href
  {\doibase 10.1063/1.2085170ÃÂ} {\bibfield  {journal} {\bibinfo
  {journal} {J. Chem. Phys.}\ }\textbf {\bibinfo {volume} {123}},\ \bibinfo
  {pages} {174101} (\bibinfo {year} {2005})}\BibitemShut {NoStop}%
\bibitem [{\citenamefont {Lee}\ \emph {et~al.}(2010)\citenamefont {Lee},
  \citenamefont {Murray}, \citenamefont {Kong}, \citenamefont {Lundqvist},\
  and\ \citenamefont {Langreth}}]{Lee2010}%
  \BibitemOpen
  \bibfield  {author} {\bibinfo {author} {\bibfnamefont {K.}~\bibnamefont
  {Lee}}, \bibinfo {author} {\bibfnamefont {{\'{E}}.}~\bibnamefont {Murray}},
  \bibinfo {author} {\bibfnamefont {L.}~\bibnamefont {Kong}}, \bibinfo {author}
  {\bibfnamefont {B.}~\bibnamefont {Lundqvist}}, \ and\ \bibinfo {author}
  {\bibfnamefont {D.}~\bibnamefont {Langreth}},\ }\href
  {papers://68dd8153-82a1-4a0e-9ae0-80ae65a14bf7/Paper/p2928} {\bibfield
  {journal} {\bibinfo  {journal} {Physical Review B}\ }\textbf {\bibinfo
  {volume} {82}},\ \bibinfo {pages} {81101} (\bibinfo {year}
  {2010})}\BibitemShut {NoStop}%
\bibitem [{\citenamefont {Azadi}\ \emph {et~al.}(2018)\citenamefont {Azadi},
  \citenamefont {Singh},\ and\ \citenamefont {K{\"{u}}hne}}]{Azadi2018}%
  \BibitemOpen
  \bibfield  {author} {\bibinfo {author} {\bibfnamefont {S.}~\bibnamefont
  {Azadi}}, \bibinfo {author} {\bibfnamefont {R.}~\bibnamefont {Singh}}, \ and\
  \bibinfo {author} {\bibfnamefont {T.~D.}\ \bibnamefont {K{\"{u}}hne}},\
  }\href {\doibase 10.1002/jcc.25104} {\bibfield  {journal} {\bibinfo
  {journal} {Journal of Computational Chemistry}\ }\textbf {\bibinfo {volume}
  {39}},\ \bibinfo {pages} {262} (\bibinfo {year} {2018})},\ \Eprint
  {http://arxiv.org/abs/1710.09703} {arXiv:1710.09703} \BibitemShut {NoStop}%
\bibitem [{\citenamefont {Li}\ \emph {et~al.}(2013)\citenamefont {Li},
  \citenamefont {Walker}, \citenamefont {Probert}, \citenamefont {Pickard},
  \citenamefont {Needs},\ and\ \citenamefont {Michaelides}}]{Li2013}%
  \BibitemOpen
  \bibfield  {author} {\bibinfo {author} {\bibfnamefont {X.-Z.}\ \bibnamefont
  {Li}}, \bibinfo {author} {\bibfnamefont {B.}~\bibnamefont {Walker}}, \bibinfo
  {author} {\bibfnamefont {M.~I.~J.}\ \bibnamefont {Probert}}, \bibinfo
  {author} {\bibfnamefont {C.~J.}\ \bibnamefont {Pickard}}, \bibinfo {author}
  {\bibfnamefont {R.~J.}\ \bibnamefont {Needs}}, \ and\ \bibinfo {author}
  {\bibfnamefont {A.}~\bibnamefont {Michaelides}},\ }\href {\doibase
  10.1088/0953-8984/25/8/085402} {\bibfield  {journal} {\bibinfo  {journal} {J.
  Phys. Condens. Matter}\ }\textbf {\bibinfo {volume} {25}},\ \bibinfo {pages}
  {085402} (\bibinfo {year} {2013})},\ \Eprint
  {http://arxiv.org/abs/arXiv:1302.0062v1} {arXiv:1302.0062v1} \BibitemShut
  {NoStop}%
\bibitem [{\citenamefont {Loubeyre}\ \emph {et~al.}(2002)\citenamefont
  {Loubeyre}, \citenamefont {Occelli},\ and\ \citenamefont
  {LeToullec}}]{Loubeyre2002}%
  \BibitemOpen
  \bibfield  {author} {\bibinfo {author} {\bibfnamefont {P.}~\bibnamefont
  {Loubeyre}}, \bibinfo {author} {\bibfnamefont {F.}~\bibnamefont {Occelli}}, \
  and\ \bibinfo {author} {\bibfnamefont {R.}~\bibnamefont {LeToullec}},\ }\href
  {papers://68dd8153-82a1-4a0e-9ae0-80ae65a14bf7/Paper/p701} {\bibfield
  {journal} {\bibinfo  {journal} {Nature}\ }\textbf {\bibinfo {volume} {416}},\
  \bibinfo {pages} {613} (\bibinfo {year} {2002})}\BibitemShut {NoStop}%
\bibitem [{Note3()}]{Note3}%
  \BibitemOpen
  \bibinfo {note} {We have re-analyzed the spectra of ref. \cite {Loubeyre2002}
  to extract the value of the indirect gap from a Tauc plot \cite {Tauc1966},
  as was performed in ref. \cite {Goncharov2013} for the data from ref. \cite
  {Howie2012b}. Details are given in the \cite {SupMat}}\BibitemShut {NoStop}%
\bibitem [{Note4()}]{Note4}%
  \BibitemOpen
  \bibinfo {note} {Our estimates of the direct gap could be biased by $\sim
  0.3eV$ due to the discreteness of our twist grid. Correcting for this bias
  will place the experimental data in between the C2/c-24 and Cmca-12
  predictions.}\BibitemShut {Stop}%
\bibitem [{Note5()}]{Note5}%
  \BibitemOpen
  \bibinfo {note} {Note that the pressure values of Ref.~\cite {Loubeyre2002}
  have been recently corrected \cite {Loubeyre2019}}\BibitemShut {NoStop}%
\bibitem [{Note6()}]{Note6}%
  \BibitemOpen
  \bibinfo {note} {This quantity is closely related to the integrated density
  of states.}\BibitemShut {Stop}%
\bibitem [{\citenamefont {Calder{\'{i}}n}\ \emph {et~al.}(2017)\citenamefont
  {Calder{\'{i}}n}, \citenamefont {Karasiev},\ and\ \citenamefont
  {Trickey}}]{Calderin2017a}%
  \BibitemOpen
  \bibfield  {author} {\bibinfo {author} {\bibfnamefont {L.}~\bibnamefont
  {Calder{\'{i}}n}}, \bibinfo {author} {\bibfnamefont {V.~V.}\ \bibnamefont
  {Karasiev}}, \ and\ \bibinfo {author} {\bibfnamefont {S.~B.}\ \bibnamefont
  {Trickey}},\ }\href {\doibase 10.1016/j.cpc.2017.08.008} {\bibfield
  {journal} {\bibinfo  {journal} {Comput. Phys. Commun.}\ }\textbf {\bibinfo
  {volume} {221}},\ \bibinfo {pages} {118} (\bibinfo {year} {2017})},\ \Eprint
  {http://arxiv.org/abs/1707.08437} {1707.08437} \BibitemShut {NoStop}%
\bibitem [{\citenamefont {Giannozzi}\ \emph {et~al.}(2017)\citenamefont
  {Giannozzi}, \citenamefont {Andreussi}, \citenamefont {Brumme}, \citenamefont
  {Bunau}, \citenamefont {{Buongiorno Nardelli}}, \citenamefont {Calandra},
  \citenamefont {Car}, \citenamefont {Cavazzoni}, \citenamefont {Ceresoli},
  \citenamefont {Cococcioni}, \citenamefont {Colonna}, \citenamefont
  {Carnimeo}, \citenamefont {{Dal Corso}}, \citenamefont {de~Gironcoli},
  \citenamefont {Delugas}, \citenamefont {DiStasio}, \citenamefont {Ferretti},
  \citenamefont {Floris}, \citenamefont {Fratesi}, \citenamefont {Fugallo},
  \citenamefont {Gebauer}, \citenamefont {Gerstmann}, \citenamefont {Giustino},
  \citenamefont {Gorni}, \citenamefont {Jia}, \citenamefont {Kawamura},
  \citenamefont {Ko}, \citenamefont {Kokalj}, \citenamefont {Kukukbenli},
  \citenamefont {Lazzeri}, \citenamefont {Marsili}, \citenamefont {Marzari},
  \citenamefont {Mauri}, \citenamefont {Nguyen}, \citenamefont {Nguyen},
  \citenamefont {Oreto-de-la Roza}, \citenamefont {Laulatto}, \citenamefont
  {Ponce}, \citenamefont {Rocca}, \citenamefont {Sabatini}, \citenamefont
  {Santra}, \citenamefont {Schlipf}, \citenamefont {Seitsonen}, \citenamefont
  {Smogunov}, \citenamefont {Timrov}, \citenamefont {Thonhauser}, \citenamefont
  {Umari}, \citenamefont {Vast}, \citenamefont {Wu},\ and\ \citenamefont
  {Baroni}}]{Giannozzi2017}%
  \BibitemOpen
  \bibfield  {author} {\bibinfo {author} {\bibfnamefont {P.}~\bibnamefont
  {Giannozzi}}, \bibinfo {author} {\bibfnamefont {O.}~\bibnamefont
  {Andreussi}}, \bibinfo {author} {\bibfnamefont {T.}~\bibnamefont {Brumme}},
  \bibinfo {author} {\bibfnamefont {O.}~\bibnamefont {Bunau}}, \bibinfo
  {author} {\bibfnamefont {M.}~\bibnamefont {{Buongiorno Nardelli}}}, \bibinfo
  {author} {\bibfnamefont {M.}~\bibnamefont {Calandra}}, \bibinfo {author}
  {\bibfnamefont {R.}~\bibnamefont {Car}}, \bibinfo {author} {\bibfnamefont
  {C.}~\bibnamefont {Cavazzoni}}, \bibinfo {author} {\bibfnamefont
  {D.}~\bibnamefont {Ceresoli}}, \bibinfo {author} {\bibfnamefont
  {M.}~\bibnamefont {Cococcioni}}, \bibinfo {author} {\bibfnamefont
  {N.}~\bibnamefont {Colonna}}, \bibinfo {author} {\bibfnamefont
  {I.}~\bibnamefont {Carnimeo}}, \bibinfo {author} {\bibfnamefont
  {A.}~\bibnamefont {{Dal Corso}}}, \bibinfo {author} {\bibfnamefont
  {S.}~\bibnamefont {de~Gironcoli}}, \bibinfo {author} {\bibfnamefont
  {P.}~\bibnamefont {Delugas}}, \bibinfo {author} {\bibfnamefont {R.~A.~J.}\
  \bibnamefont {DiStasio}}, \bibinfo {author} {\bibfnamefont {A.}~\bibnamefont
  {Ferretti}}, \bibinfo {author} {\bibfnamefont {A.}~\bibnamefont {Floris}},
  \bibinfo {author} {\bibfnamefont {G.}~\bibnamefont {Fratesi}}, \bibinfo
  {author} {\bibfnamefont {G.}~\bibnamefont {Fugallo}}, \bibinfo {author}
  {\bibfnamefont {R.}~\bibnamefont {Gebauer}}, \bibinfo {author} {\bibfnamefont
  {U.}~\bibnamefont {Gerstmann}}, \bibinfo {author} {\bibfnamefont
  {F.}~\bibnamefont {Giustino}}, \bibinfo {author} {\bibfnamefont
  {T.}~\bibnamefont {Gorni}}, \bibinfo {author} {\bibfnamefont
  {J.}~\bibnamefont {Jia}}, \bibinfo {author} {\bibfnamefont {M.}~\bibnamefont
  {Kawamura}}, \bibinfo {author} {\bibfnamefont {H.-Y.}\ \bibnamefont {Ko}},
  \bibinfo {author} {\bibfnamefont {A.}~\bibnamefont {Kokalj}}, \bibinfo
  {author} {\bibfnamefont {E.}~\bibnamefont {Kukukbenli}}, \bibinfo {author}
  {\bibfnamefont {M.}~\bibnamefont {Lazzeri}}, \bibinfo {author} {\bibfnamefont
  {M.}~\bibnamefont {Marsili}}, \bibinfo {author} {\bibfnamefont
  {N.}~\bibnamefont {Marzari}}, \bibinfo {author} {\bibfnamefont
  {F.}~\bibnamefont {Mauri}}, \bibinfo {author} {\bibfnamefont {N.~L.}\
  \bibnamefont {Nguyen}}, \bibinfo {author} {\bibfnamefont {H.-V.}\
  \bibnamefont {Nguyen}}, \bibinfo {author} {\bibfnamefont {A.}~\bibnamefont
  {Oreto-de-la Roza}}, \bibinfo {author} {\bibfnamefont {L.}~\bibnamefont
  {Laulatto}}, \bibinfo {author} {\bibfnamefont {S.}~\bibnamefont {Ponce}},
  \bibinfo {author} {\bibfnamefont {D.}~\bibnamefont {Rocca}}, \bibinfo
  {author} {\bibfnamefont {R.}~\bibnamefont {Sabatini}}, \bibinfo {author}
  {\bibfnamefont {B.}~\bibnamefont {Santra}}, \bibinfo {author} {\bibfnamefont
  {M.}~\bibnamefont {Schlipf}}, \bibinfo {author} {\bibfnamefont
  {A.}~\bibnamefont {Seitsonen}}, \bibinfo {author} {\bibfnamefont
  {A.}~\bibnamefont {Smogunov}}, \bibinfo {author} {\bibfnamefont
  {I.}~\bibnamefont {Timrov}}, \bibinfo {author} {\bibfnamefont
  {T.}~\bibnamefont {Thonhauser}}, \bibinfo {author} {\bibfnamefont
  {P.}~\bibnamefont {Umari}}, \bibinfo {author} {\bibfnamefont
  {N.}~\bibnamefont {Vast}}, \bibinfo {author} {\bibfnamefont {X.}~\bibnamefont
  {Wu}}, \ and\ \bibinfo {author} {\bibfnamefont {S.}~\bibnamefont {Baroni}},\
  }\href {\doibase 10.1088/1361-648X/aa8f79} {\bibfield  {journal} {\bibinfo
  {journal} {J. Phys. Cond Mat.}\ }\textbf {\bibinfo {volume} {29}},\ \bibinfo
  {pages} {465901} (\bibinfo {year} {2017})}\BibitemShut {NoStop}%
\bibitem [{\citenamefont {Williams}(1951)}]{Williams1951}%
  \BibitemOpen
  \bibfield  {author} {\bibinfo {author} {\bibfnamefont {F.}~\bibnamefont
  {Williams}},\ }\href
  {https://journals.aps.org/pr/pdf/10.1103/PhysRev.82.281.2} {\bibfield
  {journal} {\bibinfo  {journal} {Phys. Rev.}\ }\textbf {\bibinfo {volume}
  {82}},\ \bibinfo {pages} {281} (\bibinfo {year} {1951})}\BibitemShut
  {NoStop}%
\bibitem [{\citenamefont {Lax}(1952)}]{Lax1952}%
  \BibitemOpen
  \bibfield  {author} {\bibinfo {author} {\bibfnamefont {M.}~\bibnamefont
  {Lax}},\ }\href {\doibase 10.1063/1.1700283} {\bibfield  {journal} {\bibinfo
  {journal} {The Journal of Chemical Physics}\ }\textbf {\bibinfo {volume}
  {20}},\ \bibinfo {pages} {1752} (\bibinfo {year} {1952})}\BibitemShut
  {NoStop}%
\bibitem [{\citenamefont {Patrick}\ and\ \citenamefont
  {Giustino}(2014)}]{Patrick2014}%
  \BibitemOpen
  \bibfield  {author} {\bibinfo {author} {\bibfnamefont {C.~E.}\ \bibnamefont
  {Patrick}}\ and\ \bibinfo {author} {\bibfnamefont {F.}~\bibnamefont
  {Giustino}},\ }\href@noop {} {\bibfield  {journal} {\bibinfo  {journal}
  {Journal of Physics Condensed Matter}\ }\textbf {\bibinfo {volume} {26}},\
  \bibinfo {pages} {365503} (\bibinfo {year} {2014})}\BibitemShut {NoStop}%
\bibitem [{\citenamefont {Zacharias}\ \emph {et~al.}(2015)\citenamefont
  {Zacharias}, \citenamefont {Patrick},\ and\ \citenamefont
  {Giustino}}]{Zacharias2015}%
  \BibitemOpen
  \bibfield  {author} {\bibinfo {author} {\bibfnamefont {M.}~\bibnamefont
  {Zacharias}}, \bibinfo {author} {\bibfnamefont {C.~E.}\ \bibnamefont
  {Patrick}}, \ and\ \bibinfo {author} {\bibfnamefont {F.}~\bibnamefont
  {Giustino}},\ }\href@noop {} {\bibfield  {journal} {\bibinfo  {journal}
  {Phys. Rev. Lett.}\ }\textbf {\bibinfo {volume} {115}} (\bibinfo {year}
  {2015})}\BibitemShut {NoStop}%
\bibitem [{\citenamefont {Zacharias}\ and\ \citenamefont
  {Giustino}(2016)}]{Zacharias2016}%
  \BibitemOpen
  \bibfield  {author} {\bibinfo {author} {\bibfnamefont {M.}~\bibnamefont
  {Zacharias}}\ and\ \bibinfo {author} {\bibfnamefont {F.}~\bibnamefont
  {Giustino}},\ }\href {\doibase 10.1103/PhysRevB.94.075125} {\bibfield
  {journal} {\bibinfo  {journal} {Phys. Rev. B}\ }\textbf {\bibinfo {volume}
  {94}},\ \bibinfo {pages} {75125} (\bibinfo {year} {2016})}\BibitemShut
  {NoStop}%
\bibitem [{Note7()}]{Note7}%
  \BibitemOpen
  \bibinfo {note} {To partially correct for HSE inaccuracy, we shifted the
  energy scale by the difference between the QMC and HSE gap.}\BibitemShut
  {Stop}%
\bibitem [{\citenamefont {Pierleoni}()}]{Pierleoni2020}%
  \BibitemOpen
  \bibfield  {author} {\bibinfo {author} {\bibfnamefont {C.}~\bibnamefont
  {Pierleoni}},\ }\href@noop {} {\bibinfo  {journal} {unpublished}\
  }\BibitemShut {NoStop}%
\bibitem [{\citenamefont {Martin}\ \emph {et~al.}(2016)\citenamefont {Martin},
  \citenamefont {Reining},\ and\ \citenamefont
  {Ceperley}}]{MartinReiningCeperley}%
  \BibitemOpen
\bibfield  {journal} {  }\bibfield  {author} {\bibinfo {author} {\bibfnamefont
  {R.~M.}\ \bibnamefont {Martin}}, \bibinfo {author} {\bibfnamefont
  {L.}~\bibnamefont {Reining}}, \ and\ \bibinfo {author} {\bibfnamefont
  {D.~M.}\ \bibnamefont {Ceperley}},\ }\href@noop {} {\emph {\bibinfo {title}
  {{Interacting Electrons: Theory and Computational Approaches}}}}\ (\bibinfo
  {publisher} {Cambridge University Press},\ \bibinfo {year}
  {2016})\BibitemShut {NoStop}%
\bibitem [{\citenamefont {Perdew}(1985)}]{Perdew1985}%
  \BibitemOpen
  \bibfield  {author} {\bibinfo {author} {\bibfnamefont {J.~P.}\ \bibnamefont
  {Perdew}},\ }\href {\doibase 10.1002/qua.560280846} {\bibfield  {journal}
  {\bibinfo  {journal} {Int. J. Quantum Chem.}\ }\textbf {\bibinfo {volume}
  {28}},\ \bibinfo {pages} {497} (\bibinfo {year} {1985})}\BibitemShut
  {NoStop}%
\bibitem [{\citenamefont {Lin}\ \emph {et~al.}(2001)\citenamefont {Lin},
  \citenamefont {Zong},\ and\ \citenamefont {Ceperley}}]{Lin2001}%
  \BibitemOpen
  \bibfield  {author} {\bibinfo {author} {\bibfnamefont {C.}~\bibnamefont
  {Lin}}, \bibinfo {author} {\bibfnamefont {F.~H.}\ \bibnamefont {Zong}}, \
  and\ \bibinfo {author} {\bibfnamefont {D.~M.}\ \bibnamefont {Ceperley}},\
  }\href {\doibase 10.1103/PhysRevE.64.016702} {\bibfield  {journal} {\bibinfo
  {journal} {Phys. Rev. E}\ }\textbf {\bibinfo {volume} {64}},\ \bibinfo
  {pages} {016702} (\bibinfo {year} {2001})},\ \Eprint
  {http://arxiv.org/abs/0101339} {arXiv:0101339 [cond-mat]} \BibitemShut
  {NoStop}%
\bibitem [{\citenamefont {Holzmann}\ \emph {et~al.}(2009)\citenamefont
  {Holzmann}, \citenamefont {Bernu}, \citenamefont {Olevano}, \citenamefont
  {Martin},\ and\ \citenamefont {Ceperley}}]{Holzmann2009}%
  \BibitemOpen
  \bibfield  {author} {\bibinfo {author} {\bibfnamefont {M.}~\bibnamefont
  {Holzmann}}, \bibinfo {author} {\bibfnamefont {B.}~\bibnamefont {Bernu}},
  \bibinfo {author} {\bibfnamefont {V.}~\bibnamefont {Olevano}}, \bibinfo
  {author} {\bibfnamefont {R.~M.}\ \bibnamefont {Martin}}, \ and\ \bibinfo
  {author} {\bibfnamefont {D.~M.}\ \bibnamefont {Ceperley}},\ }\href {\doibase
  10.1103/PhysRevB.79.041308} {\bibfield  {journal} {\bibinfo  {journal} {Phys.
  Rev. B}\ }\textbf {\bibinfo {volume} {79}},\ \bibinfo {pages} {2} (\bibinfo
  {year} {2009})}\BibitemShut {NoStop}%
\bibitem [{\citenamefont {Grosso}\ and\ \citenamefont
  {Parravicini}(2014)}]{GrossoBook}%
  \BibitemOpen
  \bibfield  {author} {\bibinfo {author} {\bibfnamefont {G.}~\bibnamefont
  {Grosso}}\ and\ \bibinfo {author} {\bibfnamefont {G.}~\bibnamefont
  {Parravicini}},\ }\href@noop {} {\emph {\bibinfo {title} {Solid State
  Physics}}},\ \bibinfo {edition} {2nd}\ ed.\ (\bibinfo  {publisher} {Academic
  Press},\ \bibinfo {year} {2014})\BibitemShut {NoStop}%
\bibitem [{\citenamefont {{Wooten}}(1972)}]{WootenBook}%
  \BibitemOpen
  \bibfield  {author} {\bibinfo {author} {\bibfnamefont {F.}~\bibnamefont
  {{Wooten}}},\ }\href@noop {} {\emph {\bibinfo {title} {Optical properties of
  Solids}}}\ (\bibinfo  {publisher} {Academic Press, NY},\ \bibinfo {year}
  {1972})\BibitemShut {NoStop}%
\bibitem [{\citenamefont {Morales}\ \emph {et~al.}(2014)\citenamefont
  {Morales}, \citenamefont {Clay}, \citenamefont {Pierleoni},\ and\
  \citenamefont {Ceperley}}]{Morales2014entropy}%
  \BibitemOpen
  \bibfield  {author} {\bibinfo {author} {\bibfnamefont {M.~A.}\ \bibnamefont
  {Morales}}, \bibinfo {author} {\bibfnamefont {R.}~\bibnamefont {Clay}},
  \bibinfo {author} {\bibfnamefont {C.}~\bibnamefont {Pierleoni}}, \ and\
  \bibinfo {author} {\bibfnamefont {D.~M.}\ \bibnamefont {Ceperley}},\ }\href
  {\doibase 10.3390/e16010287} {\bibfield  {journal} {\bibinfo  {journal}
  {Entropy}\ }\textbf {\bibinfo {volume} {16}},\ \bibinfo {pages} {287}
  (\bibinfo {year} {2014})}\BibitemShut {NoStop}%
\bibitem [{\citenamefont {Tauc}\ \emph {et~al.}(1966)\citenamefont {Tauc},
  \citenamefont {Grigorovici},\ and\ \citenamefont {Vancu}}]{Tauc1966}%
  \BibitemOpen
  \bibfield  {author} {\bibinfo {author} {\bibfnamefont {J.}~\bibnamefont
  {Tauc}}, \bibinfo {author} {\bibfnamefont {R.}~\bibnamefont {Grigorovici}}, \
  and\ \bibinfo {author} {\bibfnamefont {A.}~\bibnamefont {Vancu}},\
  }\href@noop {} {\bibfield  {journal} {\bibinfo  {journal} {Physica Status
  Solidi (b)}\ }\textbf {\bibinfo {volume} {15}},\ \bibinfo {pages} {627}
  (\bibinfo {year} {1966})}\BibitemShut {NoStop}%
\end{thebibliography}%
\end{document}